\documentclass[12pt]{amsart}
\usepackage{amssymb}
\usepackage{graphics}
\usepackage{natbib}
\usepackage{bm}
\usepackage{bbm}
\usepackage{epsfig,subfigure}

\let\code=\texttt
\let\pkg=\textbf

\def\11{\mathbbm{1}}

\def\EE{\mathbb E}
\def\RR{\mathbb{R}}
\def\VV{\mathbb V}

\def\PP{\mathbb{P}}
\def\AA{\mathcal A}

\def\OO{\mathcal O}

\def\SS{\mathcal S}
\def\GG{\mathcal G}

\def\NN{\mathcal N}

\def\YY{\mathcal Y}

\newcommand{\cov}{\operatorname{\bold Cov}}

\renewcommand{\AA}{\mathcal  A}

\newcommand{\argmax}{\mbox{argmax}}

\def\sgn{\text{sgn}}

\newtheorem{example}{Example}
\newtheorem{theorem}{Theorem}

\newtheorem{definition}{Definition}
\newtheorem{proposition}{Proposition}

\renewenvironment{example}{\noindent {\bf Example.}}{\hfill{\rule{2mm}{2mm}}}

\begin{document}
\bibliographystyle{spbasic}
\title{Empirical Bayes\\ For the Reluctant Frequentist}

\author{Roger Koenker}
\author{Jiaying Gu}
\thanks{Version:  \today .}    

\begin{abstract}
    Empirical Bayes methods offer valuable tools for a large class of compound decision problems.
    In this tutorial we describe some basic principles of the empirical Bayes paradigm 
    stressing their frequentist interpretation.  Emphasis is placed on recent developments of
    nonparametric maximum likelihood methods for estimating mixture models.  A more extensive 
    introductory treatment will eventually be available in \citet{kg24}.  The methods are illustrated with 
    an extended application to models of heterogeneous income dynamics based on PSID data.
\end{abstract}
\maketitle
\pagestyle{myheadings}
\markboth{\sc Empirical Bayes}{\sc Koenker and Gu}

    \section{Introduction}
    Empirical Bayes decision theory as introduced by \citet{robbins51,robbins56}
    represented a challenge to both the \citet{wald50} and \citet{savage} strands
    of classical decision theory.  Together with the revelations of \citet{stein56}
    on the inadmissibility of the sample mean of a multivariate Gaussian vector in 
    dimensions greater than two, Robbins' results showed that compound decision problems,
    that is, ensembles of exchangeable decision problems could be fruitfully combined
    to yield improved decisions for the entire ensemble.  In effect, prior information
    could be extracted from the ensemble yielding decision rules that performed better than
    classical procedures that treated each problem in isolation.

    A simple example illustrating the benefit of this ``borrowing of strength'' from several related problems
    appears in \citet{robbins51}. Suppose we observe independent
    realizations, $Y_1, \cdots , Y_n$ with each $Y_i \sim \NN(\theta_i, 1)$ and $\theta_i \in \{ -1,1\}$.
    Our objective is to choose $\hat {\bm \theta} = (\hat \theta_1, \cdots , \hat \theta_n)$ to minimize 
    the aggregate loss,
    \[
    L(\bm{\hat\theta}, \bm \theta) = (2n)^{-1} \sum_{i=1}^n |\hat \theta_i - \theta_i |.
    \]
    The minimax rule for any single component of this problem is to choose $\hat \theta_i = \sgn(y_i)$.
    Robbins shows that this rule is also minimax for every component of the entire vector.  The least favorable
    configuration of the $\theta_i$ is one for which each component is $\pm 1$ with equal probability
    as if decided by a coin flip.  This minimax rule yields an expected loss of $\Phi(-1) \approx 0.159$;
    it is also minimax regret.  Can we do any better than this?  
    Suppose, having seen the entire sample, $y_1, \cdots , y_n$,
    we find that most of observations are positive.  Should we really stick stubbornly with the minimax view
    that the $\theta_i$ take values, $\pm 1$ with equal probability?  It seems rather perverse.
    Suppose we knew that the probability that $\theta = 1$ was $p$, then we could easily compute
    the conditional probability that $\theta = 1$ given any particular $y$ as,
     \[
    \PP_p(\theta = 1 | y) = \frac{p \varphi (y - 1)}{ p \varphi (y - 1) + (1-p) \varphi (y + 1)}.
    \]
    Evaluating the standard Gaussian density, $\varphi$, and simplifying we obtain the alternative decision rule, 
    \[
    \hat \theta_i = \sgn(y + 0.5 \log(p/(1 - p))).
    \]
    Robbins suggests that we might \emph{estimate} the probability $p$ by $\hat p = (\bar y + 1)/2$,
    and consider a plug-in version of this revised decision rule.
    The logistic adjustment has the desirable effect of biasing our estimates toward the more 
    likely of the two alternatives, for example if $\hat p = 3/4$ we would use 
    $\hat \theta_i = \sgn(y_i + .549)$, reflecting our evidence that $Y_i = 1$ was more likely
    than its alternative.  This is an empirical Bayes procedure.  There has been no reliance on 
    subjective prior information, nor on the pessimistic principle of minimaxity, only the empirical
    evidence of the observed sample has been relied upon.
    Robbins' method of moments estimator of $p$, could be replaced by other approaches such as maximum
    likelihood or even a formal conjugate Bayes procedure with a Beta prior, but ignoring the
    information contained in the ensemble of problems can be extremely costly in this setting.
    With our hypothetical $p = 3/4$ the asymptotic risk of using $\hat p$ is reduced by about
    20 percent.  This is a leading example of Robbins' claim for the ``asymptotic sub-minimaxity''
    of empirical Bayes procedures relative to the minimax procedures of Wald.

    For the Bayesophobic the foregoing example should raise no anxieties, they need  only consider
    how to go about constructing a reasonable estimate of the probability $p$.  Likewise, 
    Bayesians should be entirely comfortable computing a full posterior for $p$ that could be
    used to inform their decisions about the $\theta_i$'s.  More complicated compound decision
    problems will require more complicated estimation of the mixture structure of the problem,
    and therein lies the charm of the empirical Bayes approach.  Bayesian language and the
    machinery of Bayesian computation will prove to be convenient, but no further commitment to
    the catechism of the Reverend Bayes is required.  Indeed, we will argue that empirical
    Bayes methodology represents an alloy of the best features of the frequentist and Bayesian
    traditions.

    \section{The Compound Decision Paradigm}

    Suppose that we are faced with an exchangeable, i.e. permutation invariant, ensemble of related problems:
    \[
    Y_i \sim \varphi(y|\theta_i), \quad i = 1, 2, \cdots , n,
    \]
    where $\varphi$ denotes some familiar -- usually exponential family -- density, indexed by 
    parameters, $\theta_i$, that express the underlying heterogeneity of the problems.  
    Our task is to choose a decision rule, $\bm \delta: \YY \to \AA$, mapping realizations $Y = (Y_1, \cdots , Y_n)$
    in the sample space, $\YY$, to actions, $a = \delta (Y)$, in an action space $\AA$
    that minimize the compound risk,
    \[
    R_n (\bm \delta , \bm \theta )   = n^{-1} \EE \sum_{i=1}^n L(\bm \delta_i (Y), \theta_i).
    \]
    Following \citet{robbins56} we will restrict attention to simple, symmetric decision rules for which
    $\bm \delta_i (y_1, \cdots , y_n) = \delta (y_i)$.  Such rules are now often referred to as
    separable rules.  The function, $\delta$ may depend upon the entire
    sample, but given our objective and the exchangeable probabilistic structure of the problems there
    seems to be little merit in venturing beyond this class.  With this restriction we can write,
    \begin{align*}
	R_n (\delta , \bm \theta )   & = n^{-1} \EE \sum_{i=1}^n L(\delta (Y_i), \theta_i)\\
	 & = n^{-1} \sum_{i=1}^n \int L(\delta(y), \theta_i) \varphi(y|\theta_i)dy\\
	 & = \int \int L(\delta(y), \theta) \varphi(y|\theta)dG_n (\theta) dy
    \end{align*}
    where $G_n (\theta) = n^{-1} \sum_{i=1}^n \11 (\theta_i \leq \theta)$ denotes the empirical  
    distribution function of the $\theta_i$'s.  This is obviously an oracle risk function since we do not
    know $G_n$, but it provides a natural benchmark for evaluating the performance of various feasible decision rules.

    How related do the problems need to be?  \citet{robbins51} infamously opined,
    \begin{quote}
	No relation whatever is assumed to hold among the unknown parameters
	$\theta_i$. To emphasize this point, $Y_1$ could be an observation on a butterfly in
	Ecuador, $Y_2$ on an oyster in Maryland, $Y_3$ the temperature of a star, and
	so on, all observations being taken at different times.
    \end{quote}
    Of course, this is a bit disingenuous since we have already asserted that observations share a 
    common conditional density, and that they are exchangeable.  Ultimately, this issue of relatedness,
    or what \citet{efron10} refers to as relevance, is bound up with the form of the mixing distribution $G_n$.

    \begin{theorem}[Fundamental Theorem of Compound Decisions]  For separable decision rules compound
	risk is equal to the Bayes risk of a single copy of the compound decision problem with respect
	to the ``prior'' $G_n$.
    \end{theorem}

    Thus, an optimal decision rule for our compound decision problem can be expressed as a Bayes rule
    minimizing posterior loss with respect to the ``prior'' $G_n$:
    \begin{align*}
	B_n (\delta)  & = \int \left \{ \int_\Theta L(\delta (y), \theta) \varphi(y |\theta) dG_n (\theta) \right \} dy.\\
	& = \int \left \{ \int_\Theta L(\delta (y), \theta) h(\theta | y) d \theta \right \} f(y) dy.
    \end{align*}
    where $h(\theta | y) =  \varphi(y |\theta) dG_n (\theta)/f(y)$, is the posterior distribution of $\theta$
    given $Y = y$, and $f(y) = \int \varphi(y|\theta)dG_n (\theta)$ is the marginal density of the $Y_i$.

    From a formal Bayesian perspective exchangeability of the $Y_i$'s yields via de~Finetti's theorem 
    a mixture representation of the
    Bayes risk with a generic prior, $G$, replacing the frequentist, $G_n$.  At this point the two warring
    perspectives are essentially indistinguishable; controversy arises only when we begin to ask where can
    we find a viable $G$ that will enable us to make good decisions.  In the absence of a priori knowledge of $G$,
    this apparently requires some way to estimate the empirical distribution function
    $G_n$ of the latent parameters $\theta_i$.  In our introductory example with $\theta_i \in \{ -1,1\}$ 
    this entailed estimating a scalar probability, more general settings demand more.  We will
    consider parametric models for $G_n$ in the next section, and then proceed to consider nonparamtric
    methods in the following section.  Robbins with characteristic wit referred to this task as 
    ``Estimating the Inestimable'' in a lecture series presented at Berkeley in the late 1980s.

    \section{Parametric Priors}
    When in doubt about the form of an unknown distribution, the Gaussian shape of Napolean's hat naturally springs 
    to mind.  Suppose, instead of restricting the $\theta_i$'s to take only the values $\pm 1$, we assume instead that
    they are Gaussian with mean, zero, and variance, $\sigma_0^2$, i.e. $\theta_i \sim \NN(0,\sigma_0^2)$,
    and retain the assumption that they are embedded in standard Gaussian noise.  It follows that the
    marginal distribution of the $Y_i$'s is Gaussian with mean, zero, and variance, $1 + \sigma_0^2$.
    Under quadratic loss, $L(\delta (y), \theta) = (\delta(y) - \theta)^2$, the posterior mean, 
    \[
    \delta (y) = \EE (\theta | Y = y) = \left ( 1 - \frac{1}{1+ \sigma_0^2} \right ) y,
    \]
    is the optimal Bayes rule, provided that $\sigma_0^2$ is known.  If not, we can rely instead
    on an estimate based on $S \equiv \sum_{i=1}^n Y_i^2 \sim (1+\sigma_0^2) \chi_n^2$.  Recalling that an
    inverse $\chi_n^2$ random variable has expectation, $(n-2)^{-1}$, we obtain the method of moments,
    empirical Bayes estimator,
    \[
    \hat \delta (y) =  \left ( 1 - \frac{n-2}{S} \right ) y.
    \]
    This is the classical James-Stein estimator, \citet{jamesstein}.  It has strictly smaller compound
    risk than the inadmissible maximum likelihood estimator, $\bar \delta (y) = y$ when $n \geq 3$
    for any sequence of $\theta_i$'s.  Linear shrinkage of each coordinate toward zero may not result
    in a more accurate estimate for any one coordinate, but the compound risk of the entire vector is
    reduced.  See the Appendix for a formal exposition.

    There are many variations on the basic James-Stein rule:  if we allow the Gaussian prior to have
    (unknown) mean, $\theta_0$, estimable by the sample mean, $\bar Y_n = n^{-1} \sum_{i=1}^n Y_i$,
    we obtain the Efron-Morris rule,
    \[
    \tilde \delta (y) =  \bar Y_n + \left ( 1 - \frac{n-3}{\tilde S} \right ) (y - \bar Y_n),
    \]
    with $\tilde S \equiv \sum_{i=1}^n (Y_i - \bar Y_n)^2$, so shrinkage occurs toward the sample 
    mean rather than zero.  It can happen that $\tilde S < n-3$ in which case the shrinkage factor
    would flip the signs of the coordinate effects.  This motivates consideration of ``positive part''
    variants of the foregoing rules that restrict shrinkage to be non-negative.

    \begin{figure}
    \begin{center}
          \resizebox{ .8 \textwidth}{!}{{\includegraphics{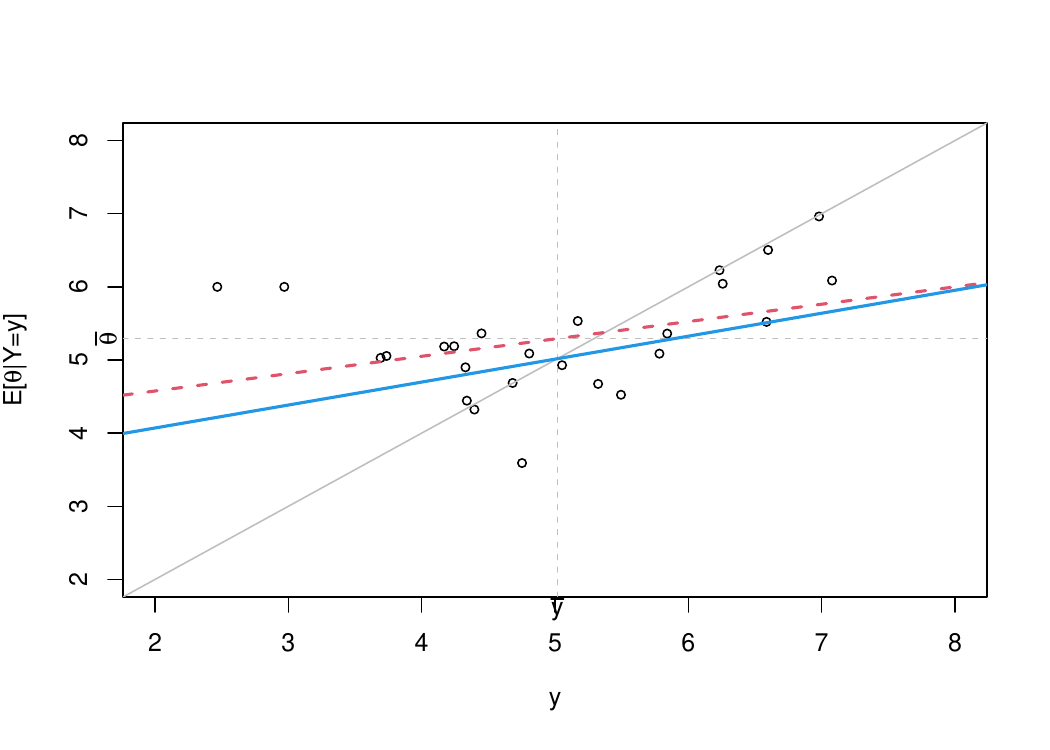}}}
	  \caption{Stigler's Galtonian perspective on Stein shrinkage views it as an attempt to
	  mimic the regression of the (latent) $\theta_i$'s on the observed $y_i$'s, noting that
	  the coefficients of this regression can be estimated without full knowledge of the
	  $\theta_i$'s using only moment informations from the $y_i$'s.  In the figure, the solid
	  line represents an oracle version of the Stein rule, while the dashed line is an estimate
	  thereof; both shrink observed $y_i$'s toward their collective sample mean in an effort to replace 
	  the naive maximum likelihood estimates, $\delta (y_i) = y_i$ (on the 45 degree line) with a more 
	  precise estimate of $\delta (y) = \EE (\theta | Y=y)$.  \label{fig.stigler}}
    \end{center} 
    \end{figure}

    Stein's revelation that the maximum likelihood estimator for the mean of a multivariate Gaussian
    random vector was inadmissible under quadratic loss in dimensions greater than two came as a surprise 
    to many practicing statisticians and is still perhaps not fully appreciated in the econometrics literature.  

    Stigler's ``Galtonian perspective on shrinkage estimators,'' \citet{stigler90}, offers a novel interpretation
    of Stein shrinkage that may aid intuition. Imagine a thought experiment in which we observe not only the $Y_i$'s, 
    but also their associated $\theta_i$'s, which we can then plot as the points appearing in Figure \ref{fig.stigler}.  
    We are interested in the conditional expectation, $\EE(\theta | Y = y)$, which is nothing but the 
    regression line in this figure.  Of course we don't have access to the actual $\theta_i$'s, so this
    might seem a bit fanciful, but the coefficients of the regression line require only the mean of the
    $\theta_i$'s and the $\cov(Y,\theta)$, which can be consistently estimated by $\bar Y_n$ and $\tilde S/(n-2) - 1$,
    respectively.  In the Figure the population regression line is depicted as the dashed red line, while the 
    sample regression is dotted and blue.  The grey 45 degree line contrasts the unshrunken, $\delta (y) = y$ estimator 
    with the two variants of the Stein rule.

    Further elaboration of the idea of modeling the prior $G$ as Gaussian may incorporate heterogeneity in
    the precision of the observed $Y_i$'s.  Consider the model,
    \[
    y_{ij} \sim \NN (\mu_i , \sigma_i^2), \quad i = 1, \dots , n, \; j = 1, \dots , J.
    \]
    The investigator observes pairs, $(\hat \mu_i , \hat \sigma_i^2 )$ with 
    $\hat \mu_i = J^{-1} \sum_j y_{ij}$ and $\hat \sigma_i^2 = (J - 1)^{-1} \sum_j (y_{ij} - \hat \mu_i)^2$;
    the objective is often to test the hypotheses:  $H_0: \mu_i = 0$, or to select a subset that violate
    these hypotheses.  In genomics this is often accomplished with a parametric empirical Bayes procedure
    implemented in the R package \pkg{limma}. 
    Rather than positing a prior on the full parameter  space, limma assumes that the $\sigma_i^2$ 
    are independent of the $\mu_i$ and are drawn iidly from the inverse chi-squared distribution,
    \[
    \sigma_i^{-2} \sim (v_0 s_0^2)^{-1} \chi_{v_0}^2 \quad i = 1, \dots , n.
    \]
    The hyperparameters, $v_0$ and $s_0$ can be estimated by maximum likelihood.  The null hypotheses
    $H_0: \mu_i = 0$ have $p$-values,
    \[
    p_i = 2 (1 - F_{t, v_0 + v} (|\tilde t_i |)),
    \]
    where $\tilde t_i = \hat \mu_i / \tilde s_i$, $\tilde s_i^2 = (v_0 s_0^2 + v \hat \sigma_i^2)/(v_0 + v)$,
    $v = J-1$ and $F_{t,v}$ is the distribution function of a Student $t$ random variable with $v$ degrees of
    freedom.  The parametric prior for the $\sigma_i^2$ shrinks the $\hat \sigma_i^2$ toward a common
    value and is intended to improve precision.  The imposition of prior information on nuisance parameters
    while leaving the parameters of primary interest, in this case the $\mu_i$, alone is referred to as
    ``partially Bayes'' by \citet{cox75}.  See \citet{is}.  More flexible nonparametric procedures will be considered
    in the next section.

    \citet{lindleysmith}, appealing to results of \citet{definetti} greatly elaborated 
    the paradigm of Gaussian linear models with parametric Gaussian priors, and thus initiated the 
    modern development of hierarchical models.  They start from the compound decision premise that 
    observations arise in an exchangeable fashion, so the $Y_i$'s come from a mixture density.
    When both the conditional density, $\varphi(y \; | \; \theta)$, and the mixing distribution, $G$, are 
    Gaussian, this leads to some elegant linear algebra.  
    \begin{proposition}[Lindley and Smith]
    Suppose, for $\theta_1 \in \RR^{p_1}$, our observed response $y \in \RR^n$ is multivariate Gaussian with 
    mean vector $A_1 \theta_1$ and covariance matrix $C_1$, i.e.  $y \sim \NN(A_1 \theta_1 , C_1)$.
    Then, if $\theta_1$ is also Gaussian, $\theta_1 \sim \NN(A_2 \theta_2 , C_2)$, the marginal distribution
    of $y$ is $\NN(A_1 A_2 \theta_2, C_1 + A_1 C_2 A_1^\top)$ and $\theta_1 | Y \sim \NN(Bb,B)$ where
    $B^{-1} = A_1^\top C_1^{-1} A_1 + C_2^{-1}$, and $b = A_1^\top C_1^{-1} y + C_2^{-1} A_2 \theta_2$. 
    \end{proposition}
    The linear model structure makes these shrinkage formulae more complicated, but it is still possible
    to recognize that the posterior mean of $\theta_1$ is a matrix weighted average of the response vector, $y$,
    and the prior mean $\theta_2$.  The closely related papers of \citet{cl1,cl2} offer further insight into this
    phenomenon.  Estimation of such hierarchical models generally requires some form of MCMC procedure, although
    when  the covariance matrices, $C_1$ and $C_2$ are spherical the formulae reduce to classical variance component
    analysis that goes back to work by \citet{balestranerlove} on the demand for natural gas in the econometrics 
    literature.

    Parametric non-Gaussian priors for linear regression models have emerged as a familiar device in modern
    high dimensional statistical modeling.  When we consider penalized regression estimators that optimize,
    \[
    \min \sum_{i=1}^n (y_i - x_i \beta)^2 + \lambda P(\beta),
    \]
    we have available a large menu of choices for the penalty function $P$.  The lasso penalty, 
    $P(\beta) = \| \beta \|_1$, employing the $\ell_1$-norm, is most familiar once we depart from the
    Gaussian ``ridge'' penalty, $P(\beta) = \| \beta \|_2^2$.  It corresponds to imposing independent Laplace 
    priors on $\beta$.  Choice of the tuning parameter, $\lambda$ controls how strongly we believe in the prior. 
    As soon as we choose $\lambda$ by cross-validation or some other form of sorcery we have ventured into the realm 
    of empirical Bayes.  

    Another prominent option for the penalty, $P$, is treat the coordinates of $\beta$
    as if they were drawn iidly from the Cauchy distribution, as considered by \citet{js04} and \citet{cvv}.  
    Although such priors are generally structured to shrink coefficients toward zero, this is typically 
    rationalized by some form of prior standardization of the design matrix.  The empirical aspect of
    these procedures is generally restricted to choice of the tuning parameter $\lambda$ representing
    the scale of the prior density.  However, more flexibility can be achieved by permitting larger
    parametric families, for example \citet{admrw} a class of Student priors for large-scale A/B testing
    settings with location, scale and degrees of freedom of the prior estimated by maximum likelihood.

\begin{figure}
    \begin{center}
          \resizebox{ .8 \textwidth}{!}{{\includegraphics{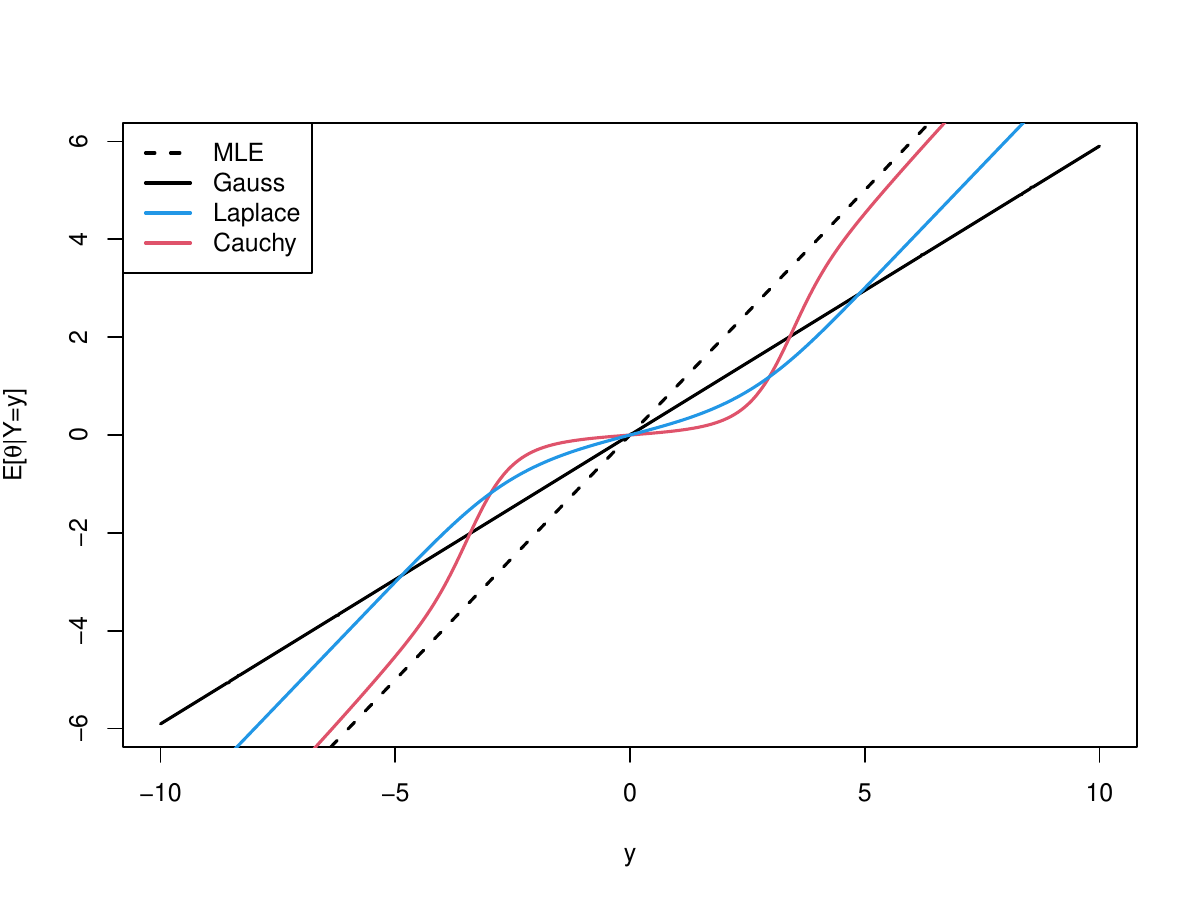}}}
	  \caption{Comparison of several parametric shrinkage methods.  \label{fig.Cauchy}}
    \end{center} 
    \end{figure}

    Returning to the simple Gaussian sequence model with scalar parameeters $\theta_i$, in Figure \ref{fig.Cauchy} 
    we contrast several forms of shrinkage: linear shrinkage with the classical Stein rule, the lasso procedure that
    shrinks moderately when $y$ is near zero, and the Cauchy penalty that shrinks very aggressively near
    zero  while large departures from zero are shrunken very little.  It should be stressed that tuning
    the location and scale of these penalties offers some flexibility, but the selection of a functional form
    for such parametric priors involves a leap of Bayesian faith that may trouble some researchers.

    Thus far we have focused entirely on settings in which our base model, $\varphi(y|\theta)$ is Gaussian.
    Parametric mixture priors play an important role in many other corners of statistics.  Poisson models
    are often paired with gamma mixing, and the modern literature on survival analysis, 
    is permeated by parametric models of ``frailty.''  As anticipated by the pioneering critique of 
    \citet{heckmansinger}, choosing a specific parametric model for frailty can be difficult so it is
    natural to turn to nonparametric methods for guidance.

    \section{Nonparametric Prior Estimation}
    Parametric models for the prior, $G$, can be difficult to choose, so it is natural to ask whether some
    nonparametric procedure can be used to estimate $G$.  An affirmative answer to this question can be
    traced back to an abstract of \citet{r50}, much more fully elaborated in \citet{kw56}.  Some
    further development of the idea of nonparametric maximum likelihood estimation of mixture models was
    made by \citet{pfanzagl88}, but practical implementation of these methods was delayed until \citet{laird}
    showed how the nascent EM algorithm, \citet{em}, could be used to compute it.  \citet{heckmansinger}
    pioneered the EM approach to explore the sensitivity to various parametric frailty models of duration
    models in econometrics.  \citet{lindsay81,lindsay} further clarified many aspects of the NPMLE, but 
    computation remained a bottleneck due to slow convergence of the EM algorithm.  Fortunately, modern
    developments in convex optimization have substantially improved computational prospects for the NPMLE.

    It is easy to see that the NPMLE problem for mixtures is convex.  We have a strictly convex objective
    subject to linear constraints over the convex set, $\GG$, of distribution functions: 
    \[
    \min_{G \in \GG} \left \{ - \sum_{i=1}^n \log f(y_i) \; | \; 
	f(y_i) = \int \varphi(y_i|\theta)dG(\theta), \; i = 1, \cdots , n \right \}
    \]
    The problem is infinite dimensional, but it can be solved to desirable precision by
    restricting the support of solutions to a finite dimensional grid $(t_1 < t_2 < \cdots < t_m)$ contained by
    the empirical support of the observations.  Finite dimensional formulations can be expressed
    given this restriction. In primal form with $A$ denoting an $n$ by $m$ matrix with typical element
    $\varphi (y_i | t_j)$, and $\bm g \in \SS_m$ denoting the masses associated with each of the grid
    points and $\SS_m$ the $m$-dimensional unit simplex, a primal formulation is given by:
    \begin{equation}
	    \min_{\bm g \in \SS_m} \left \{ - \sum_{i=1}^n \log f_i \; | \; \bm f  = A \bm g, \; \right \} \tag{P}
    \end{equation}
    and in dual form as:
    \begin{equation}
	\max_{\bm \nu \in \RR^n} \left \{\sum_{i=1}^n \log \nu_i \; | \;  A^\top \bm \nu \leq n \11_m \; \right \}.\tag{D}
    \end{equation}

    The dual form is usually somewhat more convenient for computation, and the primal solution can be
    easily recovered from the dual solution by solving for the coordinates of $\bm g$ from the linear system,
    \[
    \sum_j \varphi(y_i | t_j)g_j = \frac{1}{\hat \nu_i},
    \]
    restricted to the set $\{i : \hat \nu_i > 0 \}$, that is to those observations whose dual constraint
    is active at the dual solution.  The number of these active constraints, $m^*$, is typically far fewer than 
    its obvious upper bound of $n$, indeed it has been recently shown by \citet{pw20} that as $n \to \infty$,
    $m^* = \OO (\log n)$ when $G$ has sub-Gaussian tails.  This self-regularizing feature of the NPMLE  may come as 
    a surprise since many infinite dimensional inverse problems are ill-posed and do require some form of
    regularization, and consequent tuning parameter selection.  In contrast, the NPMLE determines of the number,
    location, and mass of the atoms of the $\hat G$ solution from the data without any interference by the analyst.

    Identifiability of $G$ in mixture models is thoroughly
    treated by \citet{teicher61,teicher67} for a scalar mixing parameter.
    \begin{definition}
    Let $\Phi(y| \theta)$ be a distribution function defined for all $\theta \in \Theta \subset \RR$ 
    and $G$ be a distribution function defined on $\Theta$, the mixture,  
    \[
    F(y) = \int_\Theta \Phi(y| \theta) dG(\theta)
    \]
    is identifiable if and only if there is a unique $G$ yielding $F$.
    \end{definition}
    For location, $\Phi (y - \theta)$, and scale, $\Phi (y\theta)$, mixtures this follows from
    the uniqueness of the characteristic function provided that the Fourier transforms of
    $\Phi(y)$ and $\Phi(e^y)$ respectively are not identically zero on some non-degenerate 
    real interval, a condition that is trivially satisfied in most of the conventional empirical
    Bayes settings in which $\varphi(y|\theta)$ is a continuous (Lebesgue) density.
    In multivariate settings and many discrete data settings identification
    conditions are more delicate and partial identification is not uncommon.  
    See \citet{kg24} for further details and references.

    Given an estimate, $\hat G$, what should we do with it?  Minimizing posterior compound loss
    is simplest with quadratic loss since it requires only computing posterior means.  
    This is particularly convenient when $\varphi$ is of the exponential family form.
    \begin{proposition} \label{prop.Tweedie} (Tweedie) For $\varphi$ of the (natural) exponential family form,
    \[
    \varphi(y|\eta) = m(y)e^{y\eta}h(\eta),
    \]
    the posterior mean is,
    \[
    \delta(y) \equiv \EE (\eta | Y = y) = \frac{d}{dy} \log ( f_G(y) / m(y)), 
    \]
    where $f_G(y) = \int \varphi(y|\eta)dG(\eta)$, is the marginal distribution of $Y$. 
    And $\delta (y)$ is non-decreasing in $y$.
    \end{proposition} 
    See Appendix \ref{app.Tweedie} for a proof and some further details.
    When $\varphi$ is standard Gaussian, so $m(y) = e^{-y^2/2}$, the posterior mean becomes, 
    \[
    \delta(y) \equiv \EE (\eta | Y = y) = y + f_G^\prime(y) / f_G (y). 
    \]
    In this Gaussian case, $f_G^\prime(y) / f_G (y)$ can be interpreted as a shrinkage
    term that pulls the naive estimator, $\hat \eta = y$ to its posterior mean.

    It is tempting to interpret the Tweedie formula as an invitation to estimate
    the marginal density $f_G$ and construct estimates of the posterior mean
    accordingly.  In the terminology of \citet{efron14}, this would be $f$-modeling
    in contrast to $g$-modeling which relies on a preliminary estimator of $G$.
    Although estimation of $f_G$ by conventional kernel methods is easy, estimation of its
    log derivative is more challenging.  More seriously, when $\varphi$ is of the
    exponential family form Proposition \ref{prop.Tweedie} asserts that the
    posterior mean function is monotone in $y$, a condition that is awkward to
    impose for standard density estimators.  See \citet{km} for a proposal for a
    shape constrained nonparametric maximum likelihood estimator of $f_G$ that
    imposes such a monotonicity constraint.

    \begin{figure}
    \begin{center}
          \resizebox{ .8 \textwidth}{!}{{\includegraphics{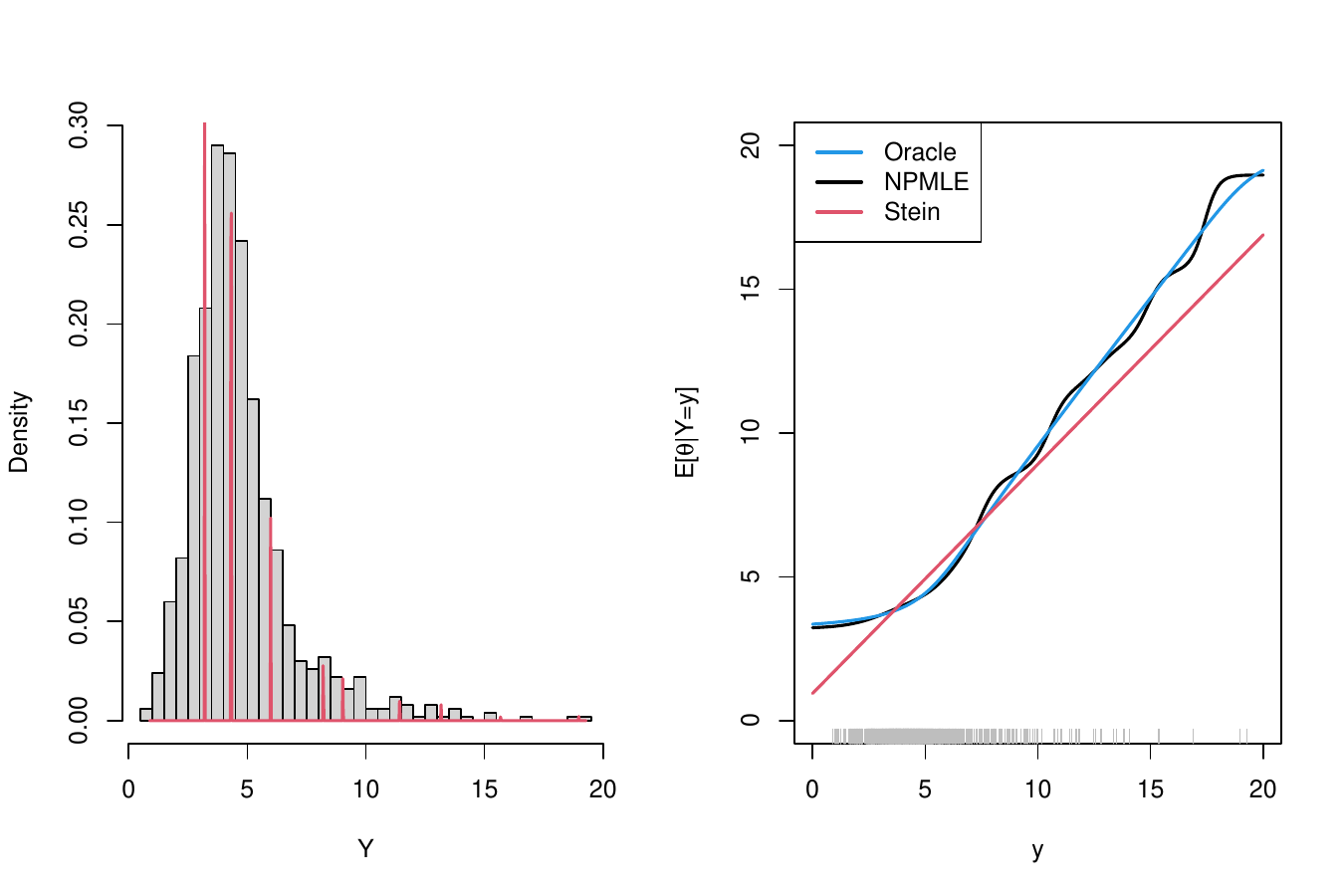}}}
	  \caption{Left panel:  Histogram of $n=1000$ observations from a standard Gaussian 
	  location mixture with 3-parameter lognormal mixing distribution.  
	  An NPLME estimate of the mixing distribution is superimposed on
	  the histogram  as the red mass points.  Right panel:  Three estimates of the 
	  posterior mean of $\theta$.
	  \label{fig.Lnorm}}
    \end{center} 
    \end{figure}

    \begin{example}
	To illustrate the NPMLE in a simple Gaussian location mixture setting, suppose that we 
	observe a random sample  of size $n = 1000$, with $Y_i \sim \NN(\theta_i, 1)$ and 
	$(\theta_i - 3)$ distributed as standard lognormal.  The left panel of Figure \ref{fig.Lnorm} 
	shows a histogram of the observed, $Y_i$'s, and superimposed in red are the mass points of
	the estimated, $\hat G$.  Of 1000 potential grid point locations for the illustrated
	NPMLE solution only 13 have associated mass in excess of 0.001.  The general shape of 
	the discrete mass points does mimic the shape of the lognormal density that was used
	to generate the sample $y_i$'s.  Viewed as a density estimate this $\hat G$ is quite
	terrible, as an estimate of a lognormal distribution function it is somewhat better.  
	But from a practical decision making perspective, $\hat G$ is useful primarily as a
	device for evaluation of smooth linear functionals like conditional means, and for 
	this it is quite excellent as noted in the Appendix.
	
	The right panel of Figure \ref{fig.Lnorm} plots posterior mean functions for three
	choices of $\hat G$.  The red line represents the Stein rule. The blue curve depicts
	the oracle rule based on full knowledge of the lognormal form of $G$.  And the black
	curve shows the estimated posterior mean corresponding to the NPMLE of $G$ depicted
	in the left panel of the figure.  The linear Stein rule clearly shrinks too much in
	both tails.  The NPMLE estimate sticks quite closely to the oracle estimate
	except in the far right tail where data is sparse as indicated by the ``rug plot''
	along the horizontal axis, and its discrete form inevitably produces some oscillation 
	around the oracle estimate.
    \end{example}

    Although the posterior mean function of the preceding example is necessarily smoothed by the 
    convolution producing $f_G$,
    and thus its log derivative, there are good reasons to consider smoother estimates of $G$
    itself.  This is especially apparent once we begin to consider inference for empirical
    Bayes point estimates or the ranking and selection problems considered in \citet{invidious}.
    A smoother alternative to the Kiefer-Wolfowitz NPMLE of $G$, proposed by \citet{efron16},
    expresses the log density of $G$ in the exponential family form,
	\[
	\log g(\theta|\alpha) = \sum_{k=1}^p z_k^\top (\theta) \alpha_k - \gamma(\alpha) 
	\]
	where $z_k (\theta) \in \RR^p, \; k = 1, \cdots , p$ are elements of a natural spline 
	basis expansion, and $\gamma$ is the
	usual constant of integration for the one-parameter exponential family density.
	The penalized Efron log likelihood is then,
	\[
	\ell (\alpha) = \sum_{i=1}^n \log f_\alpha (y_i) + c_0 \|\alpha\|,
	\]
	where the $n$ vector $f_\alpha = (f_\alpha (y_1), \cdots , f_\alpha (y_n))$ can be
	written as, $f_\alpha = AZ \alpha$ with $Z$ an $m$ by $p$ matrix with typical element
	$Z_{jk} = z_k (t_j)$. 
	Choice of the basis $Z$, its dimension $p$, and the penalty parameter $c_0$ all call for 
	some degree of expert judgement that could be guided by the relatively automatic NPMLE.
	Letting $p \to \infty$ and $c_0 \to 0$ recovers in the limit the NPMLE, although algorithms
	that fail to exploit convexity may struggle with the optimization.  An alternative
	way to achieve smoothness of an estimator of $G$ is simply to convolve the NPMLE $\hat G$
	with some smooth density.  This obviously would involve choosing  a kernel and bandwidth,
	choices that might be usefully informed by a careful examination of the discrete form of
	the NPMLE.

	When the latent parameter $\theta$ is of dimension two interior point methods for computing
	the NPMLE are still feasible using gridding as illustrated in \citet{haydn,bayesball}.  
	However, beyond dimension two such methods become unwieldy and alternative first-order 
	methods are probabily required.  Recent progress in this direction can be found in \citet{jgs},
	\citet{zcst}.

    \section{Empirical Bayes Methods for Discrete Data}
	The range of empirical Bayes methods extends far beyond the Gaussian mixture
	settings that we have emphasized thus far.  Parametric Poisson mixture models have a long
	history in actuarial risk analysis and ecology.  See, for example \citet{bs70} and \citet{corbet},
	respectively.  Given observations $y_1, \cdots , y_n$ with marginal density,
	\[
	f_G(y) =  \int \varphi ( y | \theta) dG (\theta),
	\]
	where $\PP(Y_i = y | \lambda ) = \varphi( y | \lambda_i) = e^{-\lambda_i} \lambda_i^y/y!$,
	\citet{robbins56} proposed a nonparametric estimator of the posterior mean $\EE [\theta | Y_i = y]$,
	\[
	\delta(y) = \frac{\int \theta \varphi(y|\theta) dG(\theta)}{\int \varphi(y|\theta) dG(\theta)} = 
	\frac{(y+1) f_G(y+1)}{f_G(y)}.
	\]
	Since the quantities $f_G (y)$ can be easily estimated by the observed frequencies this is an
	extremely convenient $f$-modeling strategy.  However, like other $f$-modeling methods it has
	the disadvantage that it may fail to respect the monotonicity of the Bayes rule 
	as proscribed by Proposition \ref{prop.Tweedie}.  This is particularly problematic when
	$G$ is heavy tailed since the tail frequencies of the mixture can be highly variable.
	A preferable $G$-modeling strategy is to employ the NPMLE, $\hat G$, for $G$ in the Tweedie formula
	for the posterior mean.  Although \citet{pw21} have recently shown that the original Robbins
	estimator achieves a sharp asymptotic regret bound, the NPMLE $G$-modeling approach exhibits
	significantly improved performance in simulations reported in \citet{kg24}.

	Binary response data give rise to a wide variety of mixture models that can be analysed with
	empirical Bayes methods.  In the simplest case, suppose that we have a sample $y_1, \cdots , y_n$ 
	of outcomes from binomial experiments $B(m,p_i)$ as in the tack-flipping experiment of 
	\citet{beckett94}, who describe the protocol of the experiment as follows.  
\begin{quote}
    The example involves repeated rolls of a
     common thumbtack.  A one was recorded if the tack landed point up
     and a zero was recorded if the tack landed point down.  All tacks
     started point down.  Each tack was flicked or hit with the fingers
     from where it last rested.  A fixed tack was flicked 9 times. The
     data are recorded in Table 1.  There are 320 9-tuples.  These
     arose from 16 different tacks, 2 ``flickers,'' and 10 surfaces.
     The tacks vary considerably in shape and in proportion of ones.
     The surfaces varied from rugs through tablecloths through bathroom
     floors.
\end{quote}
    Unconditionally on the type of tack and surface the experimental outcomes
    have marginal mixture density.  
	\[
	f_{G} (y) = \PP(Y = y) = \int \binom{m}{y} p^y (1-p)^{m-y}dG(p).
	\]
    Again, the mixing distribution $G$ can be estimated by maximum likelihood.
    This yields a 3-point mixture for $\hat G$ that can be reproduced by 
    running \code{demo(Bmix1)} in R from the package {REBayes}.
    This solution is essentially identical to that reported by \citet{liu96}
    using the EM algorithm although interior point convex optimization methods are
    considerably quicker.

    The binomial mixture model is easily adapted to situations with varying numbers
    of trials $m$, but a cautionary note is required regarding identification in
    such models.  Only $m+1$ distinct frequencies can be observed for $B(m,p)$
    binomials and this implies that only $m+1$ moments of $G$ are identifiable.
    Partial identification in discrete response models is discussed in more detail in 
    \citet{kg24} in the context of the \citet{klinewalters} model of employment discrimination.

    In binomial mixture models with a large number of trials it is often convenient
    to transform to the Gaussian model as for example in the extensive literature
    on baseball batting averages, see e.g. \citet{bayesball}.  In other settings
    it is more convenient to consider logistic models as for the Rasch model
    commonly used in educational testing or the Bradley-Terry model for 
    rating participants in pairwise competition.  See \citet{citations}.

    \section{Empirical Bayes Methods for Panel Data}
Longitudinal data poses many new challenges and opportunities for empirical Bayes methods.  In this section
we will reprise some prior work in \citet{haydn} on models of income dynamics and describe some
extensions that broaden applicability of such models.  The vast econometric literature on panel
data methods has gradually embraced a wider variety of latent variable formulations designed to
accommodate more general forms of heterogeneity.  The quantile autoregression framework of \citet{abb} 
is notable in this regard.  Empirical Bayes methods have a complementary role to play in this
literature and also provide a flexible approach to modeling heterogeneity in panel data.

We will begin by considering a simple Gaussian location-scale model,
\[
y_{it} = \alpha_i  + \sqrt{\theta_i} u_{it}, \quad t = 1, \cdots , m_i, 
\quad i= 1, \cdots , n
\]
with $u_{it} \sim \NN (0,1)$.  We need not interpret the $t$ subscript temporally, 
but it is frequently natural to do so. We will provisionally assume that $\alpha_i \sim G_\alpha$ and
$\theta_i \sim G_{\theta}$ are independent.  We then have sufficient statistics:
\[
\bar y_i | \alpha_i , \theta_i \sim \NN (\alpha_i, \theta_i/m_i )
\]
and
\[
S_i | r_i , \theta_i  \sim \gamma (s | r_i , \theta_i/r_i ),
\]
where $r_i = ( m_i - 1) /2$, $S_i = (m_i - 1)^{-1} \sum_{t=1}^{m_i} (y_{it} - \bar y_i)^2$,
and $\gamma(s|a,b)$ is the density of the gamma distribution with parameters, $(a,b)$.
The log likelihood becomes,
\begin{align*}
\ell (G_\alpha , G_{\theta} | y) & = K(y)\\ & + \sum_{i=1}^n   \log \int \int
\gamma(S_i | r_i , \theta /r_i ) \sqrt{m_i} 
\phi (\sqrt{m_i} (\bar y_i - \alpha_i)/\sqrt{\theta})/\sqrt{\theta} 
dG_\alpha (\alpha) dG_{\theta} (\theta).
\end{align*}
Since the scale component of the log likelihood is additively separable from the location
component, we can solve for $\hat G_{\theta}$ in a preliminary step, 
and then solve for the $\hat G_\alpha$ distribution.
In fact, under the independent prior assumption, we can  re-express the Gaussian
component of the likelihood as Student-$t$ and thereby eliminate the dependence on
$\theta$ in the NPMLE problem for estimating $G_\alpha$.   An implementation
of this estimation strategy is available with the function \code{WTLVmix} in the
R package {REBayes}.  

It is also possible to relax the independence assumption on the location and scale effects completely.
In \citet{haydn} we use longitudinal data individuals from the Panel Study on Income Dynamics (PSID)
to explore models of income dynamics with an arbitrary joint distribution of location and
scale heterogeneity.  We follow the sample selection of \citet{meghirpistaferri04} to focus on 
male head of households aged 25 - 55 with at least 9 years of consecutive earnings data. 
We further restrict  our attention to those whose earning starts from age 25 onwards. 
This leaves us with 938 individuals for whom we observe at least the early portion of their life cycle earnings. 
Among the 938 individuals, 50\% of those we observe have reported earnings of 15 years starting from age 25, 
The longest span of recorded earnings in the sample is 26 years. 

The implementation employs the function \code{WGLVmix} from the REBayes package.  
In the income dynamics application we find an apparent {\it negative}
dependence between the $\alpha$ (location) and $\theta$ scale effects indicating that low
``ability'' individuals also tend to have higher income risk.  In our prior work 
temporal dependence in the income process was specified as a simple AR(1) process whose coefficient, 
$\rho$ was estimated by profile likelihood.  

To make the AR(1) specification more explicit consider the model,
\[
\begin{array}{ll}
y_{it} & = \alpha_i + \beta_i x_{it} + v_{it} \\
v_{it} & = \rho v_{it-1} + \sqrt{\theta_i} \epsilon_{it}, \ \ \epsilon_{it} \sim \mathcal{N}(0, \sigma^{2})
\end{array}
\]
Assuming that initial conditions, $y_{i0}$, are drawn from the stationary distribution
$\NN(\alpha_i , \theta_i/ (1-\rho^2))$ is a convenient option and yields an efficient
estimator for $\rho$ provided that the assumption holds.  One could also consider less
restrictive specifications for $y_{i0}$ at the cost of introducing additional parameters
as in \citet{arellanopde}.  However it seems simpler to consider
Chamberlainian dependence of the latent effects on covariates, while trying to maintain a nonparametric
perspective, a topic we defer to future research.  As we will see, more complex short run
dynamics can be introduced via state space representations and Kalman filtering formulations of the likelihood.

Fixing $\rho$ and $\sigma^{2}$, and setting $\tilde y_{it} = y_{it} - \rho y_{it-1}$,
our model can be expressed as,
\[ 
\tilde y_{it} = (1- \rho) \alpha_i + \sqrt{\theta_i} \epsilon_{it}.
\]
And for Gaussian $\epsilon_{it}$,  sufficient statistics for 
$\alpha_i$ and $\theta_i $ are respectively the sample mean and sample variance: 
\[ 
\begin{array}{ll}
\bar y_i &= \frac{1}{m_i} \sum_{t=1}^{m_{i}} \tilde y_{it}\\
S_{i} & = \frac{1}{m_{i}-1} \sum_{t=1}^{m_{i}} (\tilde y_{it} - \bar y_i)/\sigma)^2.
\end{array}
\]
Furthermore, we have, 
$\bar y_{i} \mid \alpha_i, \theta_i  \sim 
\mathcal{N} ( (1-\rho) \alpha_i, \theta_i \sigma^2/m_i)$ and
$(m_i -1)S_i/\theta_i \mid \theta_i  \sim \chi^2_{m_i-1}$. 
Assuming the pairs $(\alpha_i, \theta_i)$  are iid with joint distribution function  
$H$, we can discretize $H$ on a two dimensional grid and write the likelihood of
observing the sample paths $(\tilde y_{i1}, \dots, \tilde y_{i,m_{i}}), \; i = 1, \cdots , n$ 
as a function of $H$,  $\rho$ and $\sigma^2$, and compute the NPMLE for the distribution $H$.  
Profile likelihood can then be employed to estimate the parameter $\rho$.

Without loss of generality,  we can set $\sigma^2 = 1$, since it is not
identified once we allow individual specific $\theta_i$ unless we were to make further moment
restrictions on $\theta_i$.  We have the following NPMLE problem:
\[
    \hat H_{\rho} := \underset{H \in \mathcal{H}}{\argmax} \prod_{i=1}^n
\int \int f(\bar y_i \mid \alpha, \theta) g(S_i \mid \theta) dH(\alpha, \theta)
\]
where $\mathcal{H}$ is the space of all bivariate distribution functions
on the domain of $\mathbb{R} \times \mathbb{R}_{+}$.  Here, $f$ is the conditional
normal density of $\bar y_i$ and $g$ is the conditional gamma density for $S_i$.
The NPMLE for $H$ is indexed by $\rho$ because both  $\bar y_i$ and $S_i$
involve $\rho$, a dependence that we have suppressed in the notation, but can be
estimated by maximizing the profile log likelihood,
\[
\ell (\rho) = 
\sum_{i=1}^n \left [ K(\bar y_i, S_i) +  \log \int \int f(\bar y_i \mid \alpha, \theta) g(S_i \mid \theta) 
d\hat H_{\rho}(\alpha, \theta) \right ].
\]

Allowing heterogeneous individual variances in earnings innovations is not new.
\citet{gewekekeane} contend that variance heterogeneity is crucial to account
for non-Gaussian features of the innovation distribution.  They use a parametric three-component 
mixture formulation.  \citet{hirano02} adopts a more flexible Dirichlet prior
specification for similar reasons.  \citet{browning10} also find significant
evidence that the variance of innovations varies across individuals.
Their model posits eight latent factors all of which are constrained to
obey parametric marginals.  They  comment ``Nowhere in the literature is there
any indication of how to specify a general joint distribution for these parameters,
nor is there any hope of identifying the joint distribution non-parametrically.''
In contrast, our approach allows only two latent factors, but has the advantage
that it does permit non-parametric estimation of their joint distribution.

\begin{figure}
    \begin{center}
	\includegraphics[width=.45\textwidth]{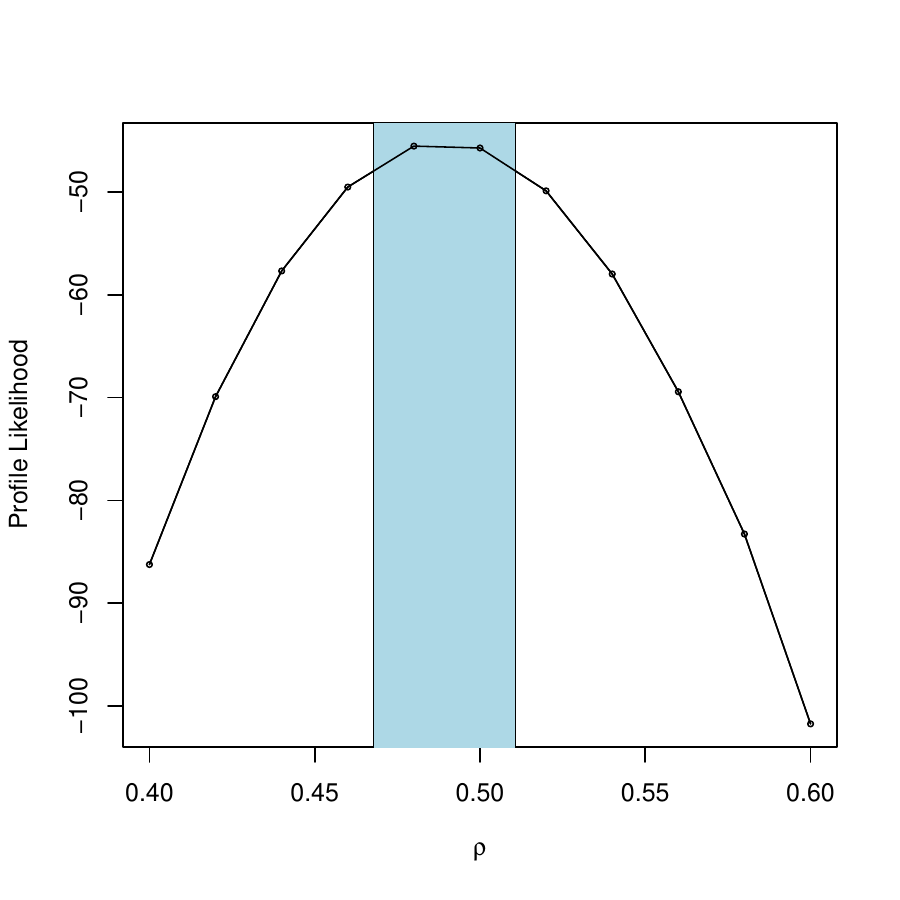}
	\includegraphics[width=.45\textwidth]{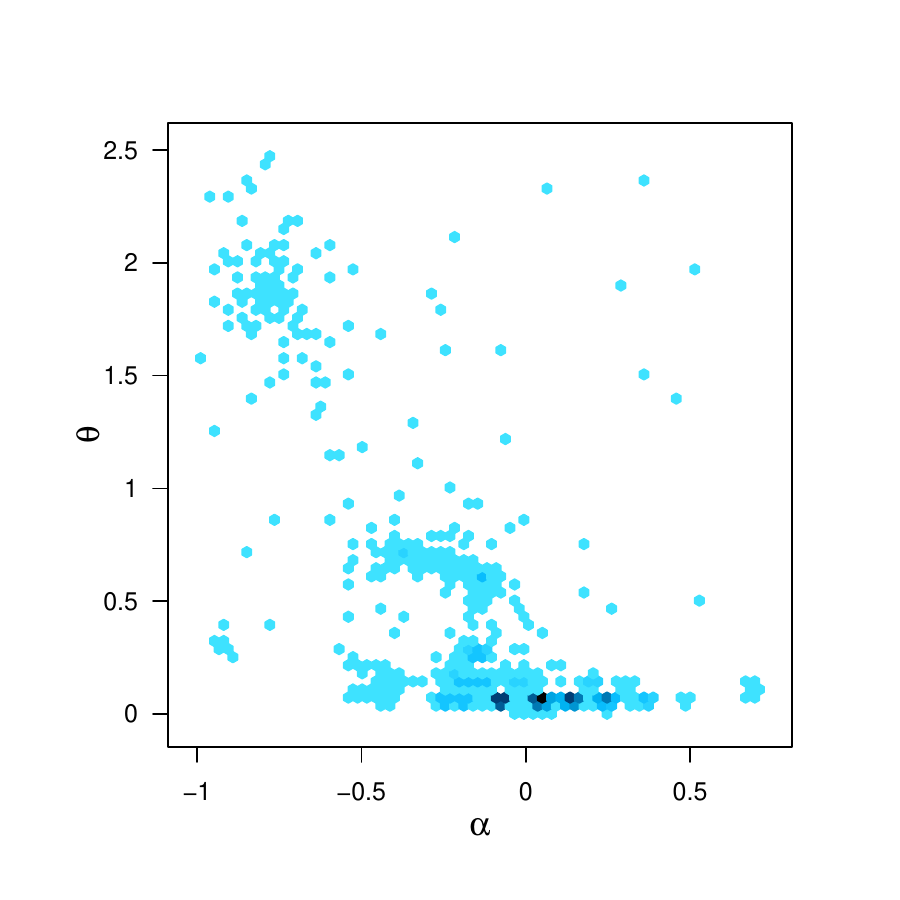}
    \end{center}
\caption{
\footnotesize{Profile Likelihood for the $\rho$ Parameter and
Heterogeneity Distribution $H(\alpha, \theta)$:
In the left panel we plot the Kiefer-Wolfowitz profile likelihood as a
function of $\rho$.  The shaded region represents a 0.95
confidence interval for $\rho$ based on a NPMLE version of the classical Wilks inversion procedure.
In the right panel we plot the estimated joint heterogeneity distribution, evaluated at the
optimal $\hat \rho$,  $\hat{H}_{\hat{\rho}}(\alpha, \theta)$. Darker hexagons
indicate greater mass, lighter ones less mass and white regions contain no mass.} }
\label{fig.Profile}
\end{figure}

The left panel of Figure \ref{fig.Profile} plots the NPMLE profile likelihood
for $\rho$, which peaks at 0.48.
The shaded region indicates a 0.95 confidence interval for $\rho$ as determined by
the classical Wilks inversion procedure, see e.g.  \citet{mvdv}, \citet{wilks}, and \citet{chenliao}.
Our estimate of $\rho$ is close to the estimate of \citet{hospido12} who also
allows an individual specific variance component in a ARCH effect variance.
She adopts a fixed effect specification for $(\alpha_i, \theta_i)$ and uses a bias
corrected estimator for $\rho$ to account for the asymptotic bias introduced by
estimating all the incidental parameters $(\alpha_i, \theta_i), i = 1, \cdots , n$.
A plausible explanation for why estimates of $\rho$ tend to be close to one
in models without heterogeneity in variances is that individual specific
variability is mistaken for AR persistence in innovations.  

The right panel of Figure \ref{fig.Profile} plots the non-parametric estimate of 
the joint distribution of $\hat H_{\hat \rho}(\alpha, \theta)$ on a $ 60 \times 60$ grid.
Mass points of the estimated distribution are indicated by shaded hexagons with
darker shading indicating more mass.  The support of  $\hat H$ is determined by
the support of the observed $(\bar y_i, S_i)$. The mixing distribution
shows some negative dependence between $\alpha$ and $\theta$, especially for
$\alpha < 0$. So low draws for $\alpha$ are more likely to be accompanied by
a more risky (higher) $\theta$.  Most of the mass of $\hat H$ is concentrated
at very low levels of $\theta$. 

\subsection{Prediction of Income Trajectories}

We would like to adapt the univariate empirical Bayes rules for prediction 
described earlier to compound decision problems for longitudinal data models.  
This objective is closely aligned with the objectives of 
\citet{chamberlainhirano99}, although our
computational methods may appear quite different. Given an initial
trajectory for an individual's earnings we would like to predict the remainder of the trajectory based 
not only on the prior history for the given individual, but also on the observed experience of 
a large sample of similar individuals.  Chamberlain and Hirano motivate this prediction exercise
as one facing a typical financial advisor.  Similar problems present themselves in many biomedical 
settings where diagnosis is based on reference growth charts or some other measures of the progression
of disease.
    
Given a trajectory $\YY_0 = \{ y_t : t = 1, \cdots , T_0 \}$ for a hypothetical individual we can
easily determine a posterior, $p (\alpha, \theta | \YY_0 )$, based on our  estimated mixture model.
These NPMLE posteriors are necessarily discrete, but we are entitled to sample from them 
for simulation purposes.  The following simulation strategy can be employed to construct an ensemble 
of completed trajectories:
\begin{enumerate}
    \item  Draw $(\alpha , \theta )$ from $p (\alpha , \theta  | \YY_0)$,
    \item  Simulate $\YY_1 = \{ y_t : t = T_0 + 1, \cdots , T \}$ as,
            $y_{T_0 + s} = \alpha + \hat \rho y_{T_0 + s - 1} + \sqrt{\theta} u_s, $ for 
            $s = 1, \cdots , T-T_0$,  and  $u_s \sim \NN (0,1)$, to obtain $m$ paths, $\YY_1$, then
    \item  Repeat steps 1 and 2, $M$ times.
\end{enumerate}
This procedure yields $mM$ trajectories from which it is easy to construct pointwise and/or uniform
prediction bands.

From a formal Bayesian perspective the foregoing  procedure may seem rather heretical.  We began with
a perfectly legitimate likelihood formulation:  data was assumed to be generated from a very 
conventional Gaussian model, except that individuals had idiosyncratic $(\alpha, \theta)$ parameters
whose joint distribution, $H$, could be viewed as a prior.  If this $H$ were delivered on a silver
platter by some local oracle we would be justified in proceeding just as we have described.  Bayes rule would allow
us to update $H$ in the light of the observed initial trajectory, $\YY_0$ for each individual,
and we would use these updated, individual specific, $\tilde H_i$'s to construct an ensemble of
forecast paths.  Various functionals of these forecast paths could then be presented.  However, lacking
a local oracle, we have relied instead on the NPMLE and our sample from PSID data to produce an $\hat H$.
Not only $H$, but also $\rho$ and potentially other model parameters are estimated by maximum
likelihood. Remarkably, no further regularization is required, and profile likelihood delivers an
asymptotically efficient estimator of these structural, i.e. ``homogeneous'' parameters.
Admittedly, we have ``sinned'' -- we've peeked at the data when we shouldn't have peeked, but our peeking has
revealed a much more plausible $H$ than we might have otherwise been expected to produce by pure introspection.
This is the charm of the empirical Bayes approach.

\begin{figure}
\begin{center}
    \resizebox{.8 \textwidth}{!}{{\includegraphics{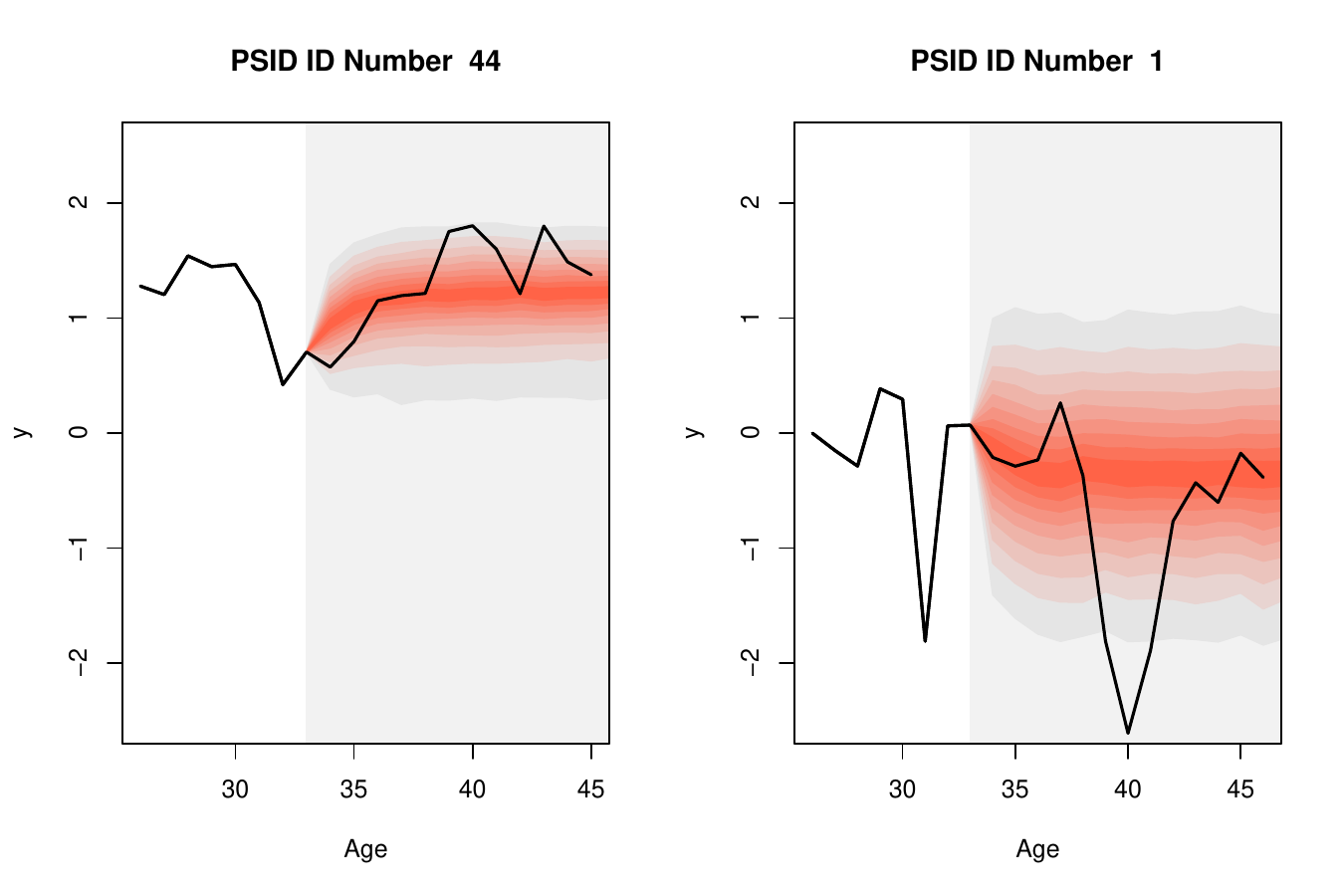}}}
\end{center}
\caption{
\footnotesize{Fan Plot of Earnings Forecasts for Two Individuals:
    Based on the initial 9 years earnings, pointwise prediction bands are shown
with graduated shading indicating bands from the 0.05 to  0.95 quantiles with 
the actual realizations superimposed as the black lines.
}}
\label{fig.pred1}
\end{figure}

\begin{figure}
\begin{center}
    \resizebox{.8 \textwidth}{!}{{\includegraphics{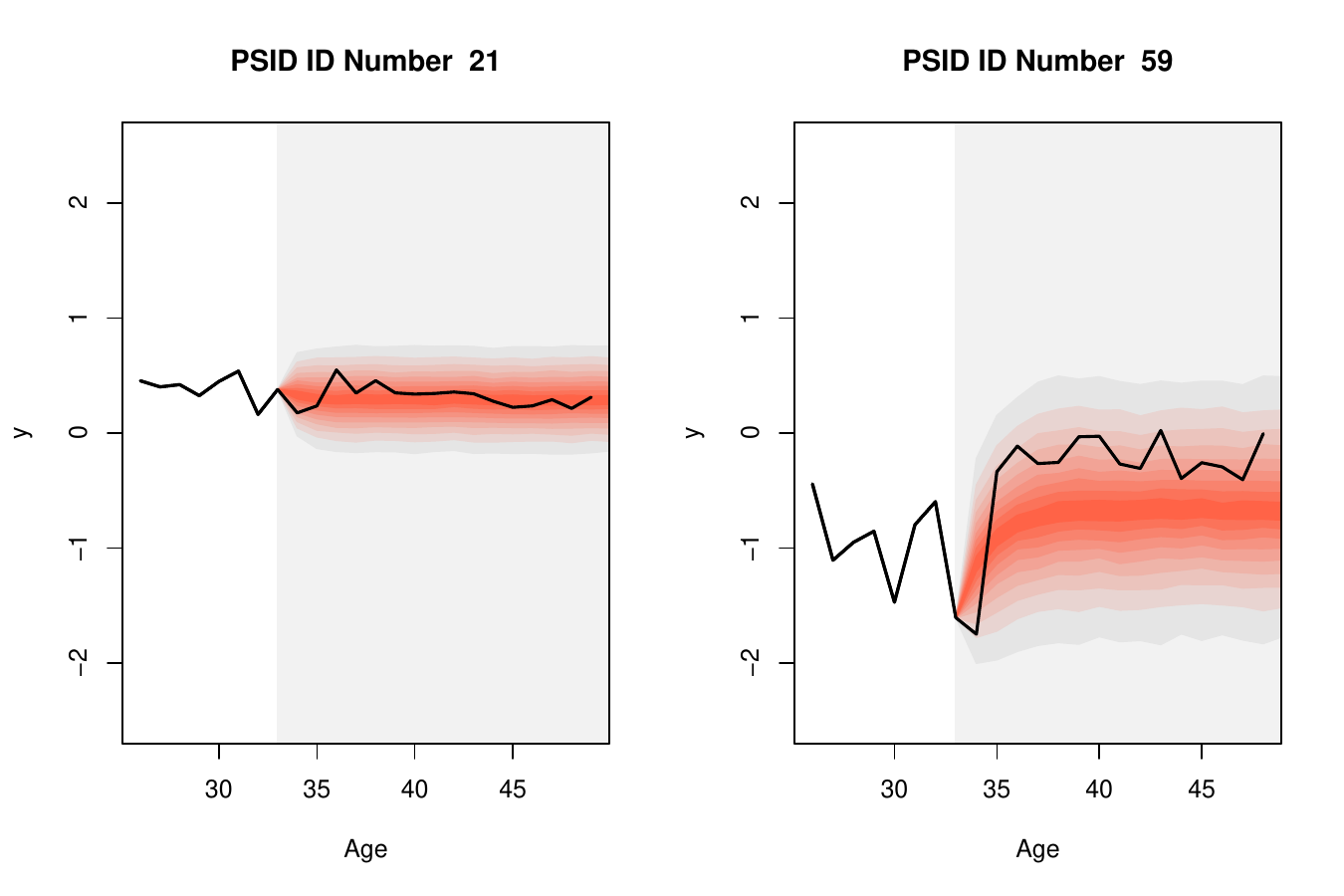}}}
\end{center}
\caption{
\footnotesize{Fan Plot of Earnings Forecasts for Two Individuals:
    Based on the initial 9 years earnings, pointwise prediction bands are shown
with graduated shading indicating bands from the 0.05 to  0.95 quantiles.
}}
\label{fig.pred2}
\end{figure}

Our prediction exercise takes $T_0 = 9$ so the first nine years of observed earnings have been used as
$\YY_0$ to construct individual specific $\tilde H_i$ that are then used to construct pointwise confidence
bands for earnings in subsequent years.  We have selected two pairs of individuals to illustrate
the variety of earnings predictions generated by our model.  In Figure \ref{fig.pred1} we contrast
predictions for an individual with relatively large mean, i.e. high $\alpha$, and large variance, high
$\theta$, with an individual with large variance, but lower mean.  The ``fan plot'' depicts
pointwise quantile prediction bands from 0.05 to 0.95 based on the simulated trajectories described above.
Realized trajectories are depicted by the solid black lines.
For the high mean individual in the left panel of Figure \ref{fig.pred1}, 
the bands are relatively narrow reflecting the fact that his  ``posterior''
assigns little mass to high $\theta$'s.  In contrast, for the lower mean individual in the right 
panel the bands are much wider, indeed the upper portion of the band overlaps with the lower portion of
the band for the higher $\alpha$ individual.  Nevertheless, we see that the lower 0.05 quantile of the
prediction band is exceeded.  Our 90\% uniform band (not shown) for this individual  just barely covers
this excursion.

In Figure \ref{fig.pred2}  we contrast high mean, low variance individual with low mean, high
variance one.  The prediction band is very narrow for the former individual in the left panel, and much wider
for the latter in the right panel.  Other features are also apparent from these figures: individuals who begin
the forecast period below their pre-forecast mean, like PSID 59, are predicted to come back
toward their mean, and some asymmetry is visible, for example in PSID 44, whose lower tail is somewhat
wider than the upper one.  Note that asymmetry requires some asymmetry in the location component of
the mixture distribution $\hat H$, since pure scale mixtures of Gaussians are necessarily
symmetric.

\subsection{Inequality and the Distribution of Annual Income Increments}

\begin{figure}
\begin{center}
    \resizebox{.8 \textwidth}{!}{{\includegraphics{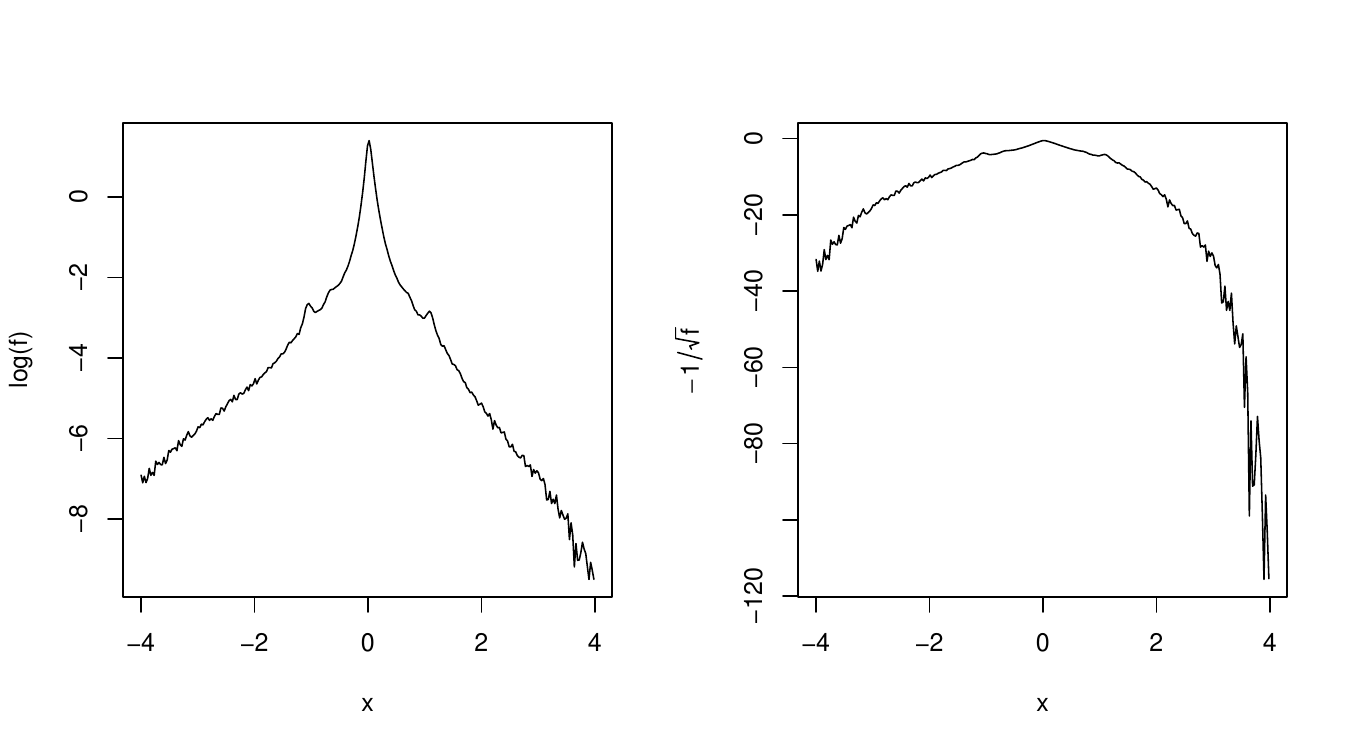}}}
\end{center}
\caption{
\footnotesize{Guvenen et al plots of annual increments of log earnings:
The left panel shows the log density plot reproduced from Figure 6 of
\citet{guvetal}, the right panel plots $-1/\sqrt{f(x)}$ yielding a much
nicer concave shape.
}}
\label{fig.Guv1}
\end{figure}

\begin{figure}
\begin{center}
    \resizebox{.8 \textwidth}{!}{{\includegraphics{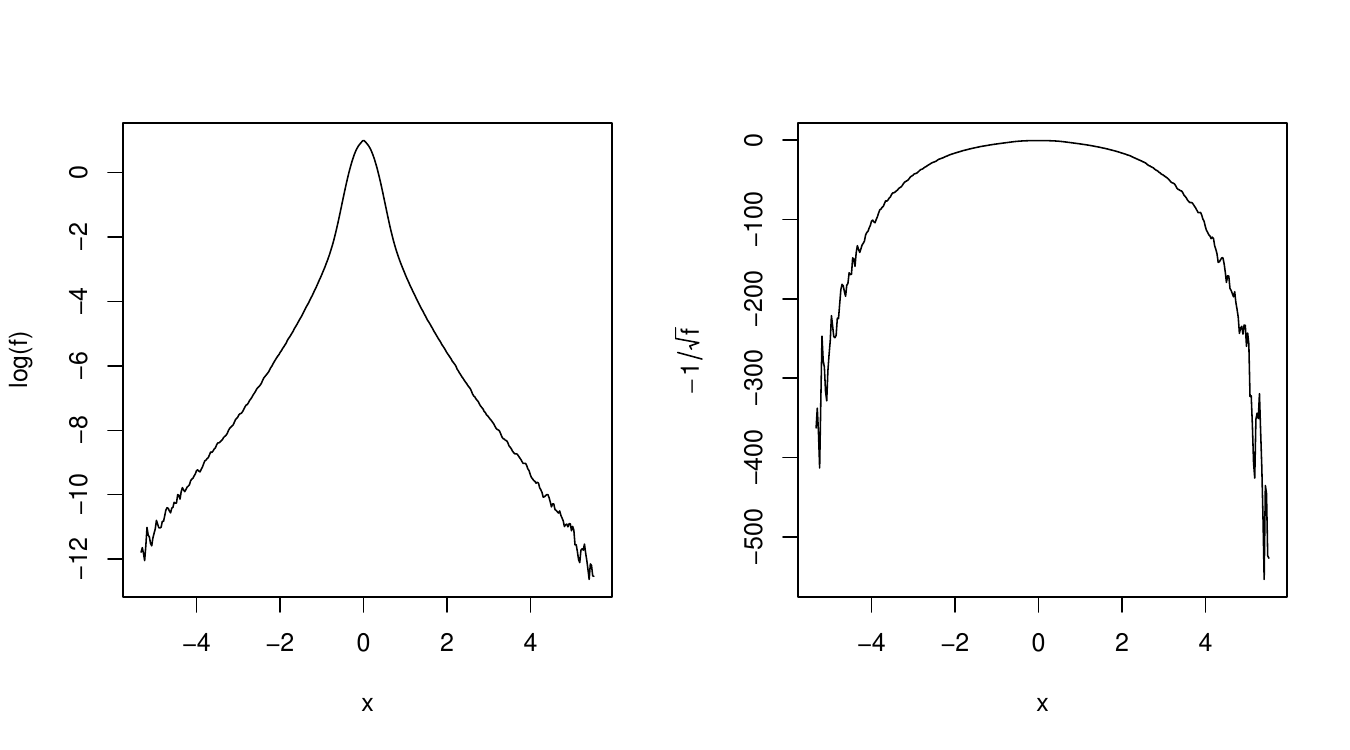}}}
\end{center}
\caption{
\footnotesize{Marginal density of annual increments of log earnings 
based on simulated data from the estimated mixture model and PSID data.
Ths simulation is based on 2500 sample paths for each of the 938 
PSID sampled individuals.  As in the previous figure, the left panel is
the log transformed density and the right panel is the $-1/\sqrt{f(x)}$
transformed density.
}}
\label{fig.sima}
\end{figure}

Inequality of incomes, wealth and other indicators of social welfare have taken
on an increasing salience as documented in the ongoing Deaton Review, \citet{deaton}.
Using a 10\% sample of U.S. Social Security records \citet{guvetal} have shown
that the distribution of annual increments in log earnings has Pareto tail behavior.
The left panel of Figure \ref{fig.Guv1} reproduces
a log-density plot appearing as their Figure 6 showing characteristic linear (Pareto) 
tail behavior with a tail exponent of about 0.40 in the left tail and 1.18 in the right
tail.  Although such densities fall outside the familiar class of log-concaves, they
are nicely accommodated by the larger class of $s$-concave densities described in
\citet{renyi}.  The right panel of Figure \ref{fig.Guv1} depicts the same density,
now plotted not as as $\log f(x)$ but as $-1/\sqrt{f(x)}$ revealing a nice concave
shape.  Such densities can be easily estimated by shape constrained non-parametric methods
as described in \citet{km.10,renyi} and \citet{hanwellner} and are implemented in the function
\code{medde} in the R package {REBayes}.

Given our estimates of the bivariate mixture model, it is of interest to see whether the estimated model
can generate a similar marginal density for annual increments in log earnings.  To investigate
this we generate 2500, $m = 50$, $M = 50$, sample paths for each of the 938 PSID sampled individuals using their
individual specific posterior distributions $\hat H_i$, and the profile likelihood point 
estimate of $\rho$.  These sample paths in log levels are then transformed to annual increments
and a marginal density for these increments is then estimated.  The resulting log and Hellinger
transformed densities are shown in Figure \ref{fig.sima}.  Not only are the shapes of the
transformed densities remarkably similar to those in the Guvenen figure, the support of the
estimated density is also remarkably consistent.  It may seem surprising that our relatively
small sample of 938 individuals from the PSID can create enough dispersion to generate this
extreme tail behavior, however further reflection suggests that the estimated scale heterogeneity
of the model is capable of generating some rather wild trajectories.

\subsection{Heterogeneous ARMA Income Dynamics}
The simple AR(1) dynamics of the preceding models is especially convenient since partial
differencing yields sufficient statistics that make the likelihood easily computed.  
However, there is a long tradition going back to \citet{friedman} of considering more
complex dynamics that decompose the income process into transitory and permanent 
components.  To illustrate how such models can be accommodated within the empirical
Bayes framework, we will consider the simple ARMA(1,1) specification adopted by 
\citet{blundellej}:
\begin{align*}
	y_{it} &= \mu_i + u_{it} + v_{it} \\
	u_{it}& = \rho u_{i,t-1} + \sigma_{\nu} \nu_{it}\\
	v_{it} & = \sigma_{\eta} \eta_{it} + \sigma_{\eta} \theta \eta_{i,t-1}.
\end{align*}
As before, $y_{it}$ denotes the residual log income process after removing covariates effects. 
For simplicity, we provisionally assume homogeneous scale parameters, $\sigma_\nu$ and $\sigma_\eta$ for
the AR and MA components, respectively.  The innovations, $\nu_{it} \sim N(0,1)$ and $\eta_{it} \sim N(0,1)$ 
are taken as iid and independent of one another.
Substituting, we have the model, 
\begin{equation} \label{eq: Blundell} 
y_{it} = \rho y_{i,t-1} + (1- \rho) \mu_i + \sigma_{\nu} \nu_{it} + 
\sigma_{\eta} \eta_{it} + (\theta - \rho) \sigma_{\eta} \eta_{i,t-1} - \rho \theta \sigma_{\eta} \eta_{i,t-2}
\end{equation}
In state-space form the model can be expressed, following \citet{harvey90} as,
\begin{align*} 
	y_{it} & = c_i + Z \alpha_{it} + G \xi_{it} \\
	\alpha_{it} & = d_i + T \alpha_{i,t-1} + H \epsilon_{it} 
\end{align*}
where 
$\alpha_{it} \in \RR^m$, 
$T \in \RR^{m \times m}$, 
$\xi_{it} \in \RR$, 
$\epsilon_{it} \in \RR^m$, 
$G \in \RR$ and $Z \in \RR^m$, 
$H \in \RR^{m \times m}$. 
The dimension of $\alpha_{it}$ is 3 in our case, $d_i = G = 0$, $c_i = \mu_i$, $Z = (1,0,0)$ and
\begin{align*}
\begin{pmatrix} \alpha_{1it} \\ \alpha_{2it} \\ \alpha_{3it}\end{pmatrix} 
    & = \begin{pmatrix} \rho & 1 & 0 \\ 0 & 0 & 1 \\ 0 & 0 & 0\end{pmatrix} 
	    \begin{pmatrix} \alpha_{1i,t-1} \\ \alpha_{2i,t-1} \\ \alpha_{3i,t-1} \end{pmatrix} 
		+ \begin{pmatrix} 1 & 1& 0 \\ \theta - \rho & 0 & 0 \\ - \theta \rho & 0 & 0 \end{pmatrix} 
		    \begin{pmatrix} \sigma_{\eta} & 0 & 0\\ 0 & \sigma_{\nu} & 0 \\ 0 & 0 & 0 \end{pmatrix} 
			\begin{pmatrix} \eta_{it} \\ \nu_{it} \\ \epsilon_{3it} \end{pmatrix}. 
\end{align*}
More explicitly, we have,
\begin{align*}
	\alpha_{1it}&=  \rho \alpha_{1i,t-1} + \alpha_{2i,t-1} + \sigma_{\eta} \eta_{it}+ \sigma_{\nu} \nu_{it}\\
	\alpha_{2it} & = \alpha_{3i,t-1} + \sigma_{\eta} (\theta - \rho)  \eta_{it} \\
	\alpha_{3it} & = - \theta \rho \sigma_{\eta} \eta_{it}, 
\end{align*}
and substituting the last two lines into the first and bring back to the first equation on $y_{it}$, we have, 
\[
y_{it} = (1- \rho) \mu_i + \rho y_{i,t-1} - \theta \rho \sigma_{\eta} \eta_{i,t-2} + 
\sigma_{\eta} (\theta - \rho) \eta_{i,t-1} + \sigma_{\eta} \eta_{it} + \sigma_{\nu} \nu_{it}
\]
which is identical to \eqref{eq: Blundell}.

Our objective is now  to estimate the latent distribution of the $\mu_i$ along with the structural 
parameters $(\rho, \theta, \sigma_{\nu}, \sigma_{\eta})$. For a time series model like \eqref{eq: Blundell}, 
the mixture likelihood of this model might appear intractable.  However, given our Gaussian assumptions
on the innovations, the likelihood -- conditional on the structural parameters -- for each trajectory
$y_{i1}, \cdots y_{iT_i}$ can be computed with the aid of the Kalman filter.  Thus, with only $\mu_i$ (location)
heterogeneity in the mixture model it is relatively straightforward to formulate the profile likelihood
problem for the structural parameters;  each entry in the NPMLE constraint matrix $A$ is supplied by the Kalman
filter, which recursively builds the likelihood evaluation for each parameter setting (see \citet{harvey90} Section 3.4). It should perhaps
be stressed that estimation of the mixing distribution for the $\mu_i$, given the structural parameters is
a convex optimization problem with a unique, quite parsimonious discrete solution.  Profile likelihood for
the remaining structural parameters is also straightforward once likelihood evaluations for the mixture
problem are in place.  In principle, there is no obstruction to reinstating scale heterogeneity into the
model, however it seemed prudent to initially consider only location heterogeneity. 

\begin{figure}
	\includegraphics[scale = 0.8]{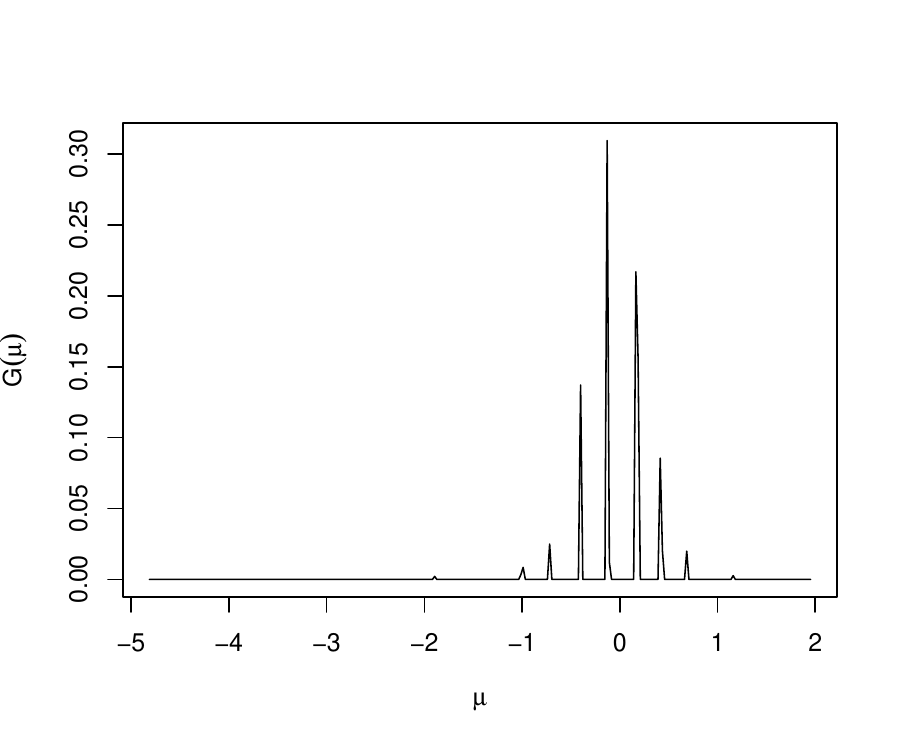}
	\caption{The estimated distribution of the individual fixed effect with the 
	heterogeneous ARMA(1,1) model.  \label{fig.Blundell}
	}
\end{figure}

Preliminary estimation of this model employing a relatively coarse grid for profiling the structural 
parameters yields, $(\hat \rho, \hat \theta, \hat \sigma_{\nu}, \hat \sigma_{\eta}) = (0.49 ,0.15, 0.17, 0.5)$. 
The corresponding estimated mixing distribution for the location parameters, $\mu_i$ is shown in 
Figure \ref{fig.Blundell}.  It was surprising to us that the persistence of the income process in
this version of the model was so weak.  In \citet{haydn} we asserted that the weak, $\rho = 0.48$,
AR(1) persistence could be attributed to the inclusion of heterogeneous scale in the model.  However, even
with fixed scale, we find similar weak persistence in the ARMA(1,1) specification implying that the
reliance on unit-root specifications of income processes may be questionable.  

We should stress, however, that we still find the heterogeneous scale specification attractive 
because it enables one to make more reliable assessments of confidence bands for posterior mean predictions.
In Appendix B we compare predictive fanplots for several representative individuals in our PSID sample. 
In the panels on the left side we have the predictions from the ARMA(1,1) model without any scale heterogeneity
while in the right panels we have the predictions from the AR(1) model with both location and scale heterogeneity.
Not unexpectedly, the ARMA(1,1) model prediction bands have the same width for all subjects, thereby over-covering
for individuals with low variability in the initial period, and under-covering for those with high variability
in the initial period.  A secondary consequence of the scale homogeneity of the ARMA(1,1) model is that it fails
to capture the extreme tail behavior illustrated in Figure \ref{fig.sima} for the AR(1) model.

\section{Conclusion}

A survey of some recent developments in empirical Bayes methods focusing on nonparametric maximum 
likelihood estimation of mixture models for latent variables has been presented.  A more extensive
development will eventually be available in \citet{kg24}.  We believe that these methods offer 
valuable new tools for studying heterogeneity in its manifold forms in economics and we look forward
to future developments.

\appendix 
    \section{Tweedie's Formula} \label{app.Tweedie}
    \citet{robbins56} attributes Proposition \ref{prop.Tweedie} to \citet{tweedie}. 
    It follows by straightforward exponential family computations, as in \citet{vhs},
    \begin{align*}
    \delta (y) & =  \EE [ \eta | Y = y]\\
    &= \int \eta \varphi(y|\eta) dG / \int  \varphi(y|\eta) dG\\
    &= \int \eta e^{y \eta} h(\eta) dG / \int e^{y \eta} h(\eta) dG \\
    &= \frac{d}{dy} \log ( \int  e^{y \eta} h(\eta) dG) \\
    &= \frac{d}{dy} \log ( f_G(y) /m(y)). 
    \end{align*}
    Differentiating again, 
    \begin{align*}
    \delta^\prime (y) & = \frac{d}{dy} 
    \left [ \frac { \int \eta \varphi dG } { \int \varphi dG } \right ]\\
    & =  \frac { \int \eta^2 \varphi dG } { \int \varphi dG } 
    - \left ( \frac { \int \eta \varphi dG } { \int  \varphi dG } \right )^2 \\
    & =  \EE [ \eta^2 | Y = y] - (\EE [ \eta | Y = y])^2\\
    & =  \VV [ \eta | Y = y] \geq 0,
    \end{align*}
    implies that $\delta$ must be monotone.  

    Stein in his discussion of \citet{efronmorris73} observed that in the standard
    Gaussian case, $Y \sim \NN(\theta, I_n)$ the oracle decision rule, $\delta(Y) = Y + \nabla \log f(Y)$, 
    under quadratic loss has compound risk,
    	\begin{align*}
	    \EE \|Y + \nabla \log f (Y)  - \theta \|^2   & = 
	    \EE \|Y - \theta \|^2 + \EE \| \nabla \log f (Y) \|^2  + 2 \EE (Y - \theta)^\top \nabla \log f(Y)\\
	    & = n + \EE \| \nabla \log f (Y) \|^2  + 2 \EE \left \{  \frac{1}{f(Y)} \nabla^2 f(Y) - 
	    \| \nabla \log f(Y) \|^2 \right \} \\
	    & = n - \EE \left \{ \| \nabla \log f (Y) \|^2  -  \frac{2}{f(Y)} \nabla^2 f(Y) \right \}\\ 
	    & = n + 4 \EE \left \{ \frac{\nabla^2 \sqrt{f(Y)}}{\sqrt{f(Y)}} \right \}
	\end{align*}
    where $\nabla$ is the vector  of first partial derivatives, and $\nabla^2$ is the Laplacian,
    $\sum \partial^2/\partial y_i^2$.  The second equality follows from Stein's lemma, and the fourth
    from the identity,
    \[
    \nabla^2 \sqrt{f} = \nabla \cdot \nabla  \sqrt{f} = \nabla \cdot \frac{\nabla f}{2\sqrt{f}} =
    \frac{1}{2\sqrt{f}} \nabla^2 f - \frac{1}{4 f^{3/2} } \| \nabla f \|^2.
    \]
    Note that the expression in the displayed equation provides an unbiased estimate of compound risk.
    Stein concludes that if $\sqrt{f}$  is superharmonic, that is, $\nabla^2 \sqrt{f} \leq 0$, 
    then the Tweedie oracle estimator, $\delta(Y) = Y + \nabla \log f (Y)$,   is minimax. 

    Of course, practical implementation of such decision rules requires an estimator for $f_G$.  
    To evaluate the cost of using an estimated decision rule, $\delta$, instead of the Tweedie oracle rule, 
    $\delta_{G}^*$, we define regret as the difference in their risks,
    \[
    \mathcal{R}_n (\delta, \GG) = \sup_{G \in \GG} \left \{ R_n (\delta, G) - R_n (\delta_G^*, G)  \right \}.
    \]
    Regret depends upon the class, $\GG$, of possible mixing distributions.  Light tailed $G$, 
    that is those with bounded support or sub-Gaussian tails, denoted $\GG_\infty$,
    make it relatively easy to estimate the marginal density.  Heavier tailed $G$, satisfying the moment
    condition, $\GG_p = \{G: \int |u|^p dG_n (u)  = \OO (1) \}$, make it more difficult.
    For estimators based on the NPMLE, $f_{\hat G}$, in the Gaussian mixture setting, \citet{jz}
    have shown that, 
    \[
    \mathcal{R}_n (\delta_{\hat G_n}, \GG) \lesssim 
    \begin{cases}
	n^{-1} (\log n)^5 & \; \mbox{if} \; \GG = \GG_\infty. \\
	n^{-p/(1+p)} (\log n)^\frac{8+9p}{2+2p} & \; \mbox{if}  \; \GG = \GG_p \; \mbox{for some fixed} \;  $p$,
    \end{cases}
    \]
    where $a_n \lesssim b_n$ denotes $a_n = \OO (b_n)$.  Theorem 1 of \citet{pw21}
    establishes that these regret bounds are minimax rate optimal up to logarithmic factors. 
    Thus, as long as $G$ is light tailed, posterior mean rules based on the NPMLE and Tweedie's formula
    achieve essentially a parametric rate of convergence up to the log factor.

\section{Predictive Distribution Comparison}
\begin{figure}
    \includegraphics[scale = 0.9]{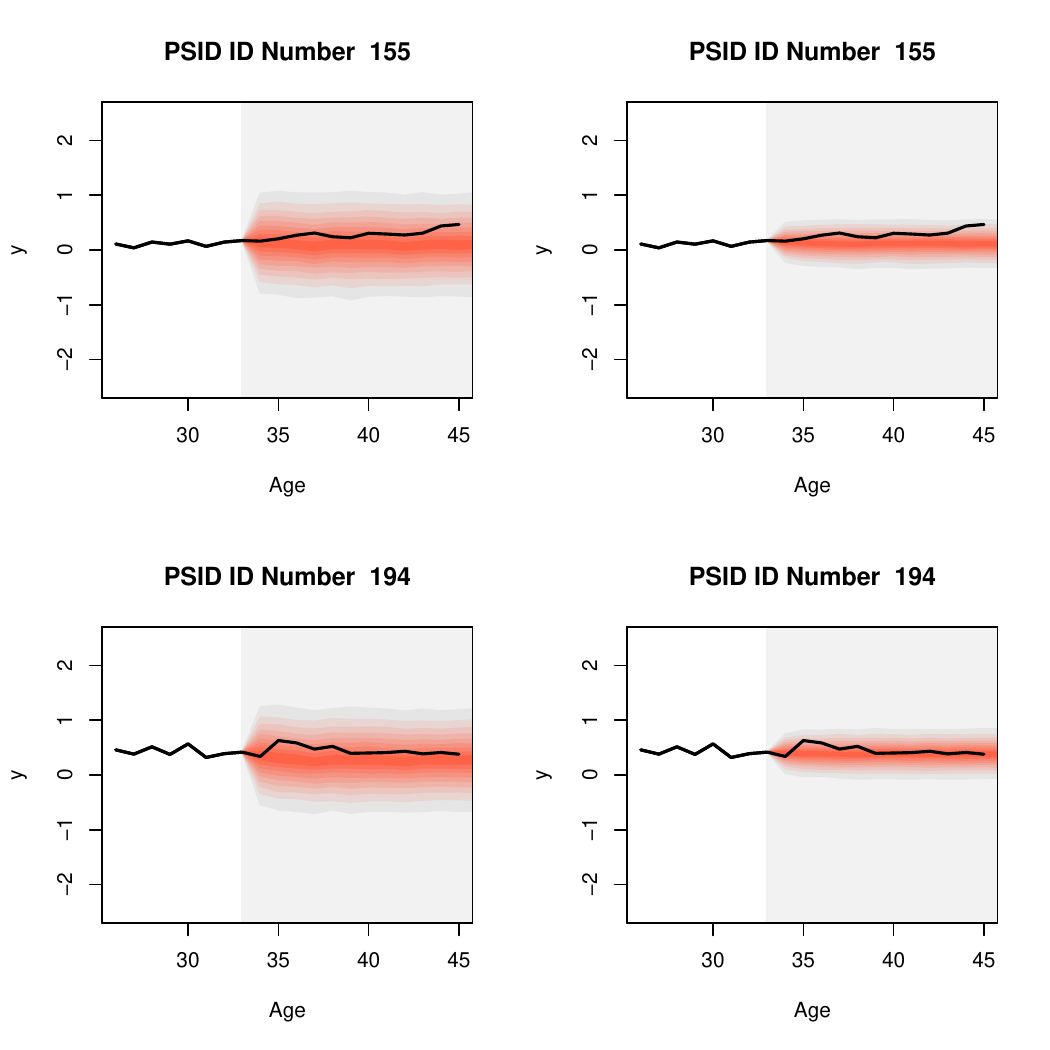}
	\caption{Left panels depict predictive bands for the ARMA(1,1) model, while
	right panels depict bands for the AR(1) heterogeneous scale model.  \label{fig.comp1}
	}
\end{figure}

\begin{figure}
    \includegraphics[scale = 0.9]{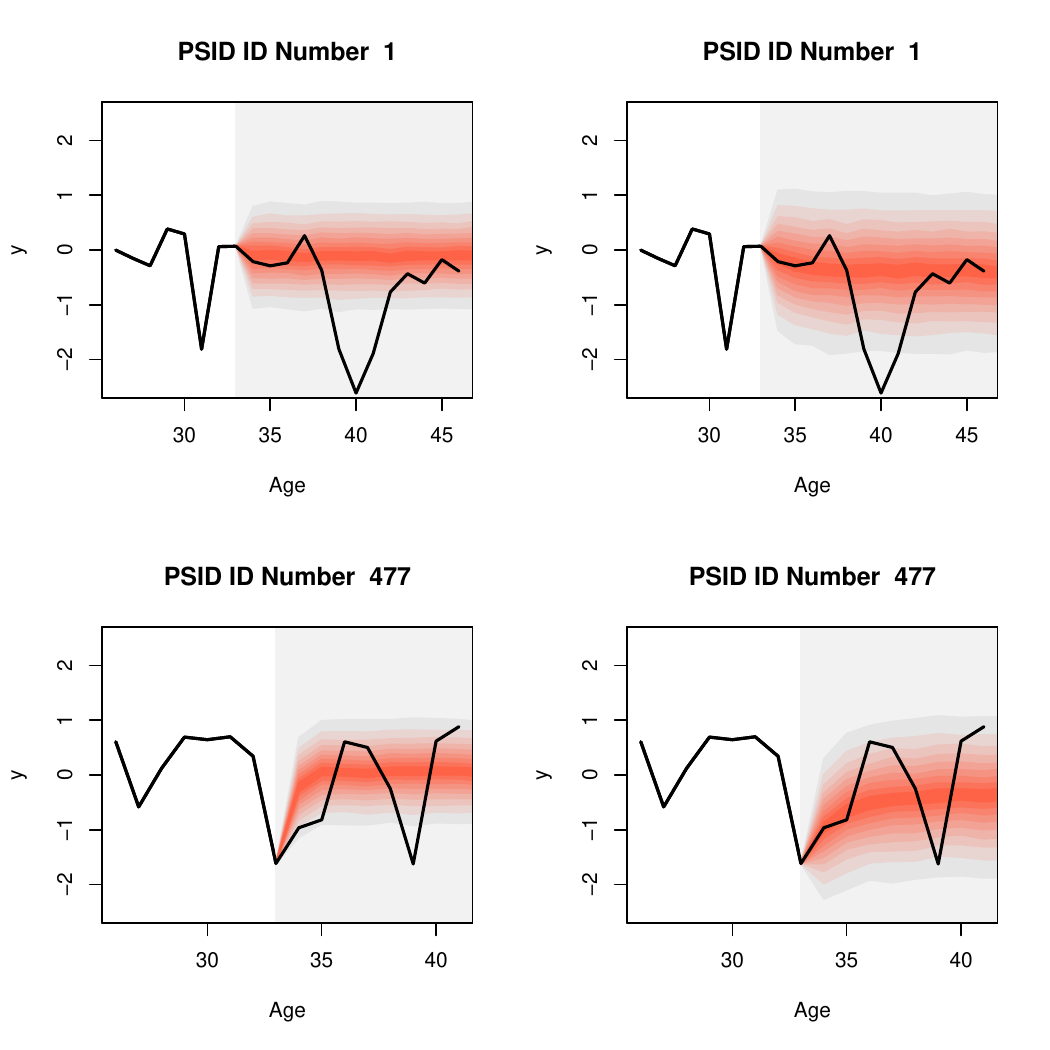}
	\caption{Left panels depict predictive bands for the ARMA(1,1) model, while
	right panels depict bands for the AR(1) heterogeneous scale model.  \label{fig.comp2}
	}
\end{figure}

In Figures \ref{fig.comp1} and \ref{fig.comp2} we compare predictive distributions 
for the scale homogeneous ARMA(1,1) model and the scale heterogeneous AR(1) model.
It can be noted that the width of the ARMA(1,1) prediction bands are the same
for both highly variable and very stable individuals in the pre-forecast period,
while the AR(1) model that incorporates individual specific scale effects adapts to this
difference.


\bibliography{EB}
\end{document}